\documentclass{jfm}
\usepackage{graphicx}
\usepackage{epstopdf, epsfig}
\usepackage{caption}
\usepackage{amsmath, amssymb}
\usepackage{mathtools}
\usepackage{subcaption}
\usepackage{soul}
\usepackage{tikz,siunitx}

\usepackage{cleveref}
\usepackage{lineno}
\usepackage{placeins}

\def\onedot{$\mathsurround0pt\ldotp$}
\def\cddot{
  \mathbin{\vcenter{\baselineskip.67ex
    \hbox{\onedot}\hbox{\onedot}}%
  }}%
\def\cdddot#1{
  \mathbin{\vcenter{\baselineskip.67ex
    \hbox{\onedot}\hbox{\onedot}\hbox{\onedot}%
  }}%
}

\shorttitle{Calculation of the mean velocity profile for strongly turbulent Taylor--Couette flow}
\shortauthor{P. Berghout et al.}

\title{Calculation of the mean velocity profile for strongly turbulent Taylor--Couette flow and arbitrary radius ratios}

\author{Pieter Berghout\aff{1}
  \corresp{\email{p.berghout@utwente.nl}},
  Roberto Verzicco\aff{1,2,3},
  Richard J. A. M. Stevens\aff{1},
  Detlef Lohse\aff{1,4},
  Daniel Chung\aff{5}}

\affiliation{\aff{1}Physics of Fluids Group and Max Planck Center Twente, MESA+ Institute and J. M. Burgers Centre for Fluid Dynamics, University of Twente, P.O. Box 217, 7500AE Enschede, Netherlands
\aff{2}Dipartimento di Ingegneria Industriale, University of Rome `Tor Vergata', Via del Politecnico 1, Roma 00133, Italy
\aff{3}Gran Sasso Science Institute -- Viale F. Crispi, 7 67100 L'Aquila, Italy.
\aff{4}Max Planck Institute for Dynamics and Self-Organisation, Am Fassberg 17, 37077 G\"{o}ttingen, Germany
\aff{5}Department of Mechanical Engineering, University of Melbourne, Victoria 3010, Australia
}

\begin{document}

\maketitle

\begin{abstract}
Taylor--Couette (TC) flow is the shear-driven flow between two coaxial independently rotating cylinders. In recent years, high-fidelity simulations and experiments revealed the shape of the streamwise and angular velocity profiles up to very high Reynolds numbers. However, due to curvature effects, so far no theory has been able to correctly describe the turbulent streamwise velocity profile for given radius ratio, as the classical Prandtl--von K\'arm\'an logarithmic law for turbulent boundary layers over a flat surface at most fits in a limited spatial region.

Here we address this deficiency by applying the idea of a Monin--Obukhov curvature length to turbulent TC flow. This length separates the flow regions where the production of turbulent kinetic energy is governed by pure shear from that where it acts in combination with the curvature of the streamlines. We demonstrate that for all Reynolds numbers and radius ratios, the mean streamwise and angular velocity profiles collapse according to this separation. We then derive the functional form of the velocity profile. Finally, we match the newly derived angular velocity profile with the constant angular momentum profile at the height of the boundary layer, to obtain the dependence of the torque on the Reynolds number, or, in other words, of the generalized Nusselt number (i.e., the dimensionless angular velocity transport) on the Taylor number. 
\end{abstract}

\section{Introduction}
Most flows in nature and engineering are bounded by solid walls. In general, the flow in the immediate vicinity -- at a molecular scale distance -- from the wall has the velocity of the wall, the so-called no-slip boundary condition. As a consequence, a steep gradient in the mean streamwise velocity profiles exists within the boundary layer (BL) region between the wall and the freely flowing fluid above. In the BL, the action of viscosity against the gradient of the streamwise velocity results in viscous dissipation, the conversion of kinetic energy into heat. 

\subsection{Turbulent flow over a flat plate: Prandtl--von K\'arman BL theory}
For slowly flowing fluids (low Reynolds numbers), the edge of the BL remains smooth, and the fluid flow in the BL is two-dimensional. This laminar BL is described by the famous Prandtl--Blasius self-similar solution \citep{sch79}. However, for fast flowing fluids (high Reynolds numbers), the BL becomes turbulent, and the flow inside the BL becomes vortical and three-dimensional. Although exact solutions of these turbulent BLs do not exist, a well-established functional form of the mean streamwise velocity can be obtained based on simple dimensional arguments \citep{sch79}. 
The hallmark result therefrom can be obtained from realizing that the mean streamwise velocity gradient in the wall-normal direction ($\frac{du}{dy}$) is a function of two dimensionless parameters only \citep{pop00},
\begin{equation}
\label{eq:vgrad}
\frac{du}{dy} = \frac{u_\tau}{y}\Phi \left( \frac{y}{\delta_\nu},\frac{y}{\delta} \right),
\end{equation} 
where $u_\tau$ is the friction velocity defined as $u_\tau=\sqrt[•]{\tau_w/\rho}$, $\tau_w$ is the mean wall shear stress, $\rho$ is the fluid density, $\delta$ is the outer length scale (e.g. the BL thickness), and $\delta_\nu$ is the viscous length scale $\delta_\nu=\nu/u_\tau$, with $\nu$ the kinematic viscosity of the fluid. Non-dimensionalization by the viscous scales $u_\tau$ and $\delta_\nu$ is indicated by a superscript `+'. We define the friction Reynolds number based on these viscous quantities as $Re_\tau = \frac{u_{\tau,i}d}{2\nu}$, where $d$ is the gap width between the two rotating cylinders. The yet undefined function $\Phi(\frac{y}{\delta_\nu},\frac{y}{\delta})$ must go to a constant ($=\kappa^{-1}$) when $\delta_\nu \ll y \ll \delta$, which is known as the inertial sublayer. In this limit, we can integrate (\ref{eq:vgrad}) and arrive at the celebrated logarithmic law of the wall for turbulent BLs over a flat surface: 
\begin{equation}
\label{eq:karmanlaw}
u^+ = \kappa^{-1} \log{y^+} + B.
\end{equation}
This law is connected with the names of Prandtl and von K\'arm\'an. It is supported by overwhelming experimental and numerical evidence (e.g. \cite{smi11}). The values of the two parameters are $\kappa\approx 0.4$ and $B\approx 5.0$.

An important extension of the theory concerns buoyancy stratified BLs, where an additional forcing acts on the wall-normal momentum component. A prominent example of such a system is the atmospheric surface layer, where thermal forcing stabilizes or destabilizes the flow. The thermal stratification introduces, aside from $\delta_\nu$ and $\delta$, a third relevant length scale: the Obukhov length $L_{ob}$ \citep{obu71}. This length $L_{ob}$ is proportional to the distance from the wall above which the production of turbulence is significantly affected by buoyancy, and below which the production of turbulence is governed purely by shear. With the introduction of this length $L_{ob}$, (\ref{eq:vgrad}) becomes:
\begin{equation}
\label{eq:vgrad_L}
\frac{du}{dy} = \frac{u_\tau}{y}\Phi \left(\frac{y}{\delta_\nu},\frac{y}{\delta}, \frac{y}{L_{ob}} \right),
\end{equation}
which was first proposed by \citep{mon54}. For the inertial sublayer ($\delta_\nu \ll y \ll \delta$) only the dependence on $\frac{y}{L_{ob}}$ remains. Various empirical fits exist for $\Phi(\frac{y}{L_{ob}})$. Evidently, in the limit of $\frac{y}{L_{ob}}\ll 1$ they must obey $\Phi(\frac{y}{L_{ob}})=\kappa^{-1}$, thus indicating that buoyancy plays no role. We point to \S 4 of \cite{mon75} for an in-depth analysis of stratified BLs.

\subsection{Turbulent flow with streamwise curvature: Taylor--Couette turbulence}
Whereas flat plate BLs are often studied, and the existence of a logarithmic profile of the mean streamwise velocity is well established, the study of flows with streamwise curvature is less developed, despite its ubiquity, e.g. ship hulls or turbomachinery. In this paper, we attempt to narrow this gap. One canonical system for flow in a curved geometry is Taylor--Couette (TC) flow. 
TC flow is the shear-driven flow in between two coaxial, independently rotating cylinders. Since the physical system is closed, one can derive a global balance between the differential rotation of the cylinders and the total energy dissipation in the flow, which is directly related to the torque ($T$) on any of the cylinders \citep{gro16}. 

The dimensionless torque $G$ is defined as $G\equiv T/(\rho\nu^2L_z)$, where $L_z$ is the height of the cylinder. It depends on the Reynolds number of the inner and outer cylinder, defined as $Re_{i,o}=\omega_{i,o}r_{i,o}d/\nu$. Here, $r_{i,o}$ is the radius of the inner (outer) cylinder, $\omega_{i,o}$ is the angular velocity of the inner (outer) cylinder, $d$ is the gap width, and $\nu$ is the kinematic viscosity. The relation $G(Re_i, Re_o)$ is directly connected to the structure of the mean velocity profile. Uncovering this relation -- for its fundamental implications and practical relevance -- can be considered the primary research question. 

In this paper we consider pure inner cylinder rotation ($Re_o=0$), for which in the laminar case \cite{tay23b} derived that $G\propto Re$. For intermediate $Re$, \cite{mar84} -- in analogy to the work of \cite{mal58} on Rayleigh--B\'enard (RB) flow -- argued by exploring marginal stability arguments that $G\propto Re^{5/3}$. He modelled the flow domain as being partitioned into a turbulent bulk region with constant angular momentum L \citep{tow56} and two laminar BLs. For high but finite $Re$, the BLs become turbulent \citep{gro12, ost15b, kru17}, and the effective scaling exponent increases with increasing $Re$ \citep{lat92a, lat92}. Analogous to the interpretation of the strongly turbulent regime by \cite{kra62} and \cite{cha97} in RB flow, \cite{gro11} derived logarithmic corrections to the $G(Re)$ scaling, coming from the turbulent BLs, such that $G\propto Re^2 \times \text{log(Re)-corrections}$. In the limit of $Re\to \infty$, the dissipation will anywhere in the flow scale with the velocity difference cubed, irrespective of the length scale \citep{lat92a}, resulting in a torque scaling of $G\propto Re^2$.

High-fidelity data on the structure of the BL are essential for testing all proposed scaling relationships. Therefore, much work has been carried out to determine the mean streamwise velocity profile at high $Re$. \cite{hui12} used particle image velocimetry (PIV) and laser doppler velocimetry (LDA) to study the turbulent BL at an unprecedented resolution. For $\eta=0.716$, where $\eta$ is the radius ratio, they find that for high $Re_i$, i.e. $Re_i=O(10^6)$, the classical logarithmic BL exists only in a very limited spatial region of $50 < y^+ < 600$. \cite{vee16} employed PIV to study the velocity profiles at low radius ratio of $\eta=0.50$, for which the curvature effects are stronger, and find no von K\'arm\'an type logarithmic BL. For $\eta=0.91$, \cite{ost14} and \cite{ost15b} employed direct numerical simulations (DNSs) and find that the slope of the mean streamwise velocity profile is ever changing with $Re_i$, at least up to $Re_i=O(10^5)$. We further note that \cite{gro14} argue that the appropriate velocity that obeys the classical von K\'arm\'an profile is the angular velocity, rather than the streamwise velocity, based on conservation laws of the Navier--Stokes equations in this axial symmetry.

In this paper we will explain that the introduction of a curvature length scale delineates the region where one can expect a shear-dominated turbulent BL and another region where curvature effects will alter the structure of the flow, similar as the Obukhov length in stratified shear flow separates the shear dominated regime from the buoyancy dominated regime. This paper is organized as follows: In \S 2 we will give the Navier--Stokes equations and boundary conditions for TC flow. In \S 3 we will discuss the used datasets. We will then, in \S \ref{sec:res1}, derive a functional form for the angular velocity throughout the entire BL for arbitrary Reynolds numbers but only for pure inner cylinder (IC) rotation. We extend the theory towards varying radius ratios in \S \ref{sec:res4}. Finally, we match the BL and bulk velocity profiles and arrive at a new functional form for Nu(Ta) and $\text{C}_\text{f}(\text{Re}_i)$ for TC in \S \ref{sec:res5}. The paper ends with conclusions and an outlook.

\section{Navier--Stokes equations for Taylor--Couette flow}
\label{sec:tc}
\begin{figure}
\centering
\includegraphics[width=0.50\linewidth]{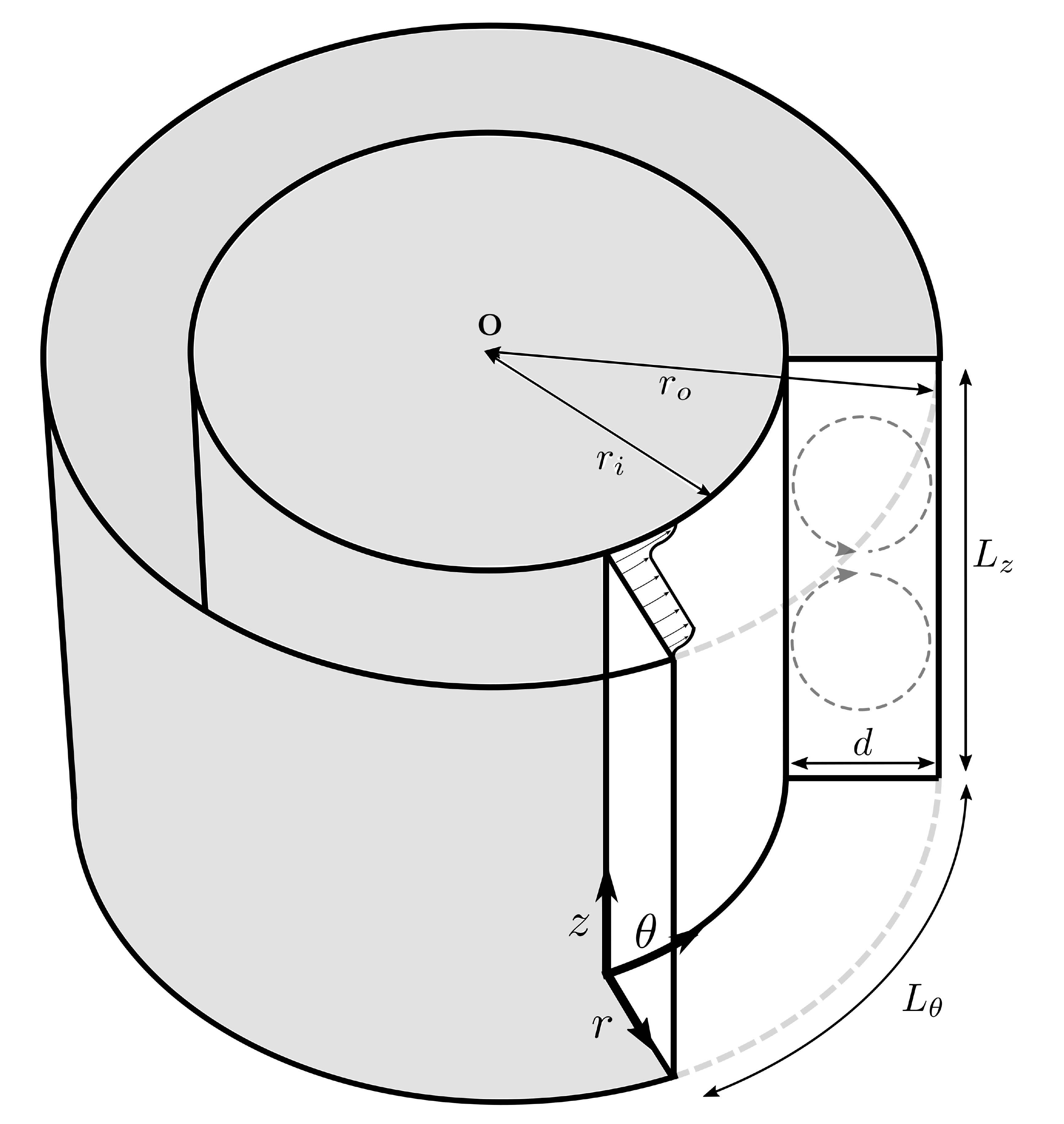}
\caption{Schematic of TC flow including the coordinate directions $(\theta, z, r)$, IC radius $r_i$, OC radius $r_o$, gap width $d$, the spanwise (axial) extent of the flow domain $L_z$ and the streamwise extent of the flow domain $L_\theta$, which is used in DNSs that employ periodic boundary conditions in the azimuthal directions. $\eta=r_i/r_o$ is the radius ratio. The grey dashed arrows represent the turbulent Taylor vortices. }
\label{fig:tc}
\end{figure}
When the inner cylinder rotates and the outer cylinder (OC) remains stationary (the case to which we restrict us in this paper), TC flow is linearly unstable \citep{ray16b}. The ratio between the destabilizing centrifugal force and the stabilizing viscous force is expressed by the Taylor number \citep{tay23b},
\begin{equation}
\text{Ta} = \frac{(1+\eta)^4}{64\eta^2}\frac{(r_o-r_i)^2(r_i+r_o)^2(\omega_i-\omega_o)^2}{\nu^2}.
\end{equation}
The Reynolds number $\text{Re}_{i,o}$ is related to Ta via the relation $\text{Re}_i-\eta\text{Re}_o=\frac{\text{Ta}^{1/2}}{f(\eta)}$ with $f(\eta) = \frac{(1+\eta)^3}{8\eta^2}$. \cite{eck07b} showed that the mean angular velocity flux 
\begin{equation}
J^\omega = r^3 \large[\langle u_r \omega \rangle_{A(r),t} - \nu \partial_r  \langle \omega \rangle_{A(r),t}\large]
\end{equation}
is independent of $r$, where $\langle . \rangle_{A(r),t}$ refers to averaging over a cylindrical surface $A(r)$ and time $t$. The torque $T$ per unit length is related to $J^\omega$ by $T = 2\pi \rho J^\omega$. Therefore also $T$ is constant with $r$. 

TC flow, see the schematic in figure \ref{fig:tc}, is described by the three components of the Navier--Stokes equations in an inertial frame in cylindrical coordinates as \citep{ll87}, with $w_r$ the radial velocity, $u_\theta$ the azimuthal velocity and $v_z$ the axial velocity
\begin{equation}
\label{eq:ns1}
\partial_t w_r + (\textbf{u}\cdot \nabla)w_r -\frac{u_\theta^2}{r} = -\partial_r P_t + \nu \left \{\mathcal{4}
 w_r - \frac{2}{r^2}\partial_\theta u_\theta - \frac{w_r}{r^2}  \right \},
\end{equation}
\begin{equation}
\label{eq:ns2}
\partial_t u_\theta + (\textbf{u}\cdot \nabla)u_\theta + \frac{w_r u_\theta}{r} = -\frac{1}{r}\partial_\theta P_t + \nu \left \{\mathcal{4}
 u_\theta + \frac{2}{r^2}\partial_\theta w_r - \frac{u_\theta}{r^2}  \right \},
\end{equation}
\begin{equation}
\label{eq:ns3}
\partial_tv_z + (\textbf{u}\cdot \nabla)v_z = -\partial_z P_t + \nu \mathcal{4}
 v_z,
\end{equation}
where the operators are, 
\begin{equation}
\label{eq:grad1}
(\textbf{u}\cdot \nabla)f = w_r\partial_r f + \frac{u_\theta}{r}\partial_\theta f + v_z\partial_z f,
\end{equation}
and
\begin{equation}
\label{eq:grad2}
\mathcal{4}f = \frac{1}{r}\partial_r (r\partial_r f) + \frac{1}{r^2}\partial_\theta^2 f + \partial_z^2 f,
\end{equation}
with for IC rotation only, the boundary conditions $w_r(r_i)=w_r(r_o)=0$, $v_z(r_i)=v_z(r_o)=0$, $u_\theta(r_i)=r_i\omega_i$ and $u_\theta(r_o)=r_o\omega_o=0$.
Note that $P_t$ is the kinematic pressure, and $\rho P_t$ is the physical pressure. The continuity equation reads 
\begin{equation}
\frac{1}{r}\partial_r(rw_r) + \frac{1}{r}\partial_\theta u_\theta + \partial_z v_z =0.
\end{equation}

\section{Employed datasets}\label{sec:dataset}
In this paper we apply our theoretical analysis to published datasets with varying radius ratio, see Table \ref{table_l} in the appendix. We now briefly describe the techniques that are used to acquire these datasets. However, we refer to the original papers for more details. 

\cite{hui12} did experiments on highly turbulent inner cylinder rotating TC flow with the Twente turbulent TC facility ($T^3C$) \citep{gil11a}, with the radius ratio $\eta=0.716$ and the aspect ratio $\Gamma=11.7$. In particular, they carried out PIV and particle tracking velocimetry (PTV) to measure the mean and the variance of the streamwise velocity profiles at $9.9\times10^8 \le \text{Ta} \le 6.2\times10^{12}$, for both the IC BL and the OC BL. 

\cite{vee16} performed experiments on turbulent TC flow in the classical turbulent regime (i.e., before the BLs become turbulent) with the Cottbus TC facility \citep{mer13}, with radius ratio $\eta=0.50$ and aspect ratio $\Gamma=20$. They carried out PIV to measure the mean streamwise and wall-normal velocity profiles at $5.8\times10^7 \le \text{Ta} \le 6.2\times10^{9}$. Although \cite{vee16} carried out both counter rotation and pure inner cylinder rotation experiments, we will discuss here the latter dataset only. 

\cite{ost15b} carried out DNSs of highly turbulent IC rotating TC flow by using a second-order finite-difference scheme \citep{ver96, poe15c}. With a radius ratio of $\eta=0.909$ they simulated three cases with $1.1\times10^{10} \le \text{Ta} \le 1.0\times 10^{11}$. Additionally, they simulated a large gap case, $\eta=0.5$, with $\text{Ta}=1.1\times10^{11}$. For all cases the aspect ratio was fixed at $\Gamma=2\pi/3$.

\section{Velocity profiles in Taylor--Couette turbulence}
\label{sec:res1}
Whereas effects of spanwise curvature on the profiles were investigated before \citep{gro17}, in this section we set out to develop a new functional form of the mean angular velocity profile $\omega^+(y^+)$ (with $\omega^+ = \omega/\omega_\tau$ and $\omega_{\tau,(i,o)}=u_{\tau,(i,o)}/r_{(i,o)}$) in that part of the IC BL and OC BL where the streamwise curvature effects are significant. Note that (\ref{eq:vgrad}) can also be postulated for $\omega(y)$, so that the gradient becomes
\begin{equation}
\label{eq:omgrad}
\frac{d\omega}{dy} = \frac{\omega_\tau}{y}\Phi_\omega \left( \frac{y}{\delta_\nu},\frac{y}{\delta} \right),
\end{equation}
where $\Phi_\omega \left( \frac{y}{\delta_\nu},\frac{y}{\delta} \right)$ goes to a constant in the inertial region $\delta_\nu \ll y \ll \delta$. 
We follow the conclusion of \cite{gro14}, namely that near the wall the \textit{angular} velocity $\omega^+(y^+)$ fits to a logarithmic form closer than the \textit{azimuthal} velocity $u^+(y^+)$, and we apply our analysis to $\omega^+(y^+)$. For reference we have added figure \ref{fig:u_lc} in the appendix, where we apply the analysis (see following pages) to the \textit{azimuthal} velocity profile. A slightly less convincing collapse of the azimuthal velocity profiles, in comparison to the angular velocity profiles, indicates that the angular velocity profile is indeed the appropriate quantity.  

In \S \ref{ssec:derLc} we first derive the curvature Obukhov length and then apply our analysis to the highest Re dataset available \citep{hui12}. Subsequently, we analyse both the IC BL (\S \ref{ssec:ICBL}) and OC BL (\S \ref{ssec:OCBL}) and in \S \ref{ssec:BULK} also the constant angular momentum region in the bulk.

\subsection{Derivation of the curvature Obukhov length $L_c$}
\label{ssec:derLc}
\begin{figure}
\centering
\includegraphics[width=0.6\linewidth]{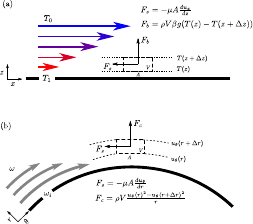}
\caption{A schematic representation of the analogy between the effects of buoyancy and streamline curvature on a BL. (a) A flat plate unstably stratified BL. The volume element, with volume $V$, top and bottom surface area $A$, and height $\Delta z$, exemplifies the working of the shear force $F_s$ and the buoyancy force $F_b$. Note that $\beta$ is the thermal expansion coefficient and $g$ is the gravitational acceleration that is defined positive in the $-z$ direction. (b) A top view of a BL over a curved surface (e.g. the TC IC). In analogy to $F_b$ in (a), the centrifugal force $F_c$ works in the wall normal direction, and in the case of IC rotation, destabilizes the flow. }
\label{fig:ill_ana}
\end{figure}

Following \cite{bra69}, we draw the analogy between the effects of buoyancy and streamline curvature on turbulent shear flow. Therefore it is informative to assess the balance of turbulent kinetic energy (TKE) in the flow. To do so, we first Reynolds-decompose the velocity and pressure field ((\ref{eq:ns1}) to (\ref{eq:ns3})), such that $\mathbf{v} = \mathbf{U} + \mathbf{u}$, where $\mathbf{v}=(w_r, u_\theta, v_z)$ is the full velocity, $\mathbf{U}=(W,U,V)$ is the time averaged velocity and $\mathbf{u}= (w,u,v)$ is the fluctuating component. Upon multiplying the decomposed Navier--Stokes equations by $\mathbf{u}$, and then taking the time average, we arrive at the TKE equations. In vector notation, with the definition of TKE (strictly speaking the turbulent intensity since we divide by $\rho$) being $q = \frac{1}{2}(\overline{u^2} + \overline{v^2} + \overline{w^2})$, the TKE equation reads (see also \cite{mos84}):
\begin{equation}
\label{eq:tke}
\begin{gathered}
\partial_t q + \nabla\cdot (q\mathbf{U}) + \frac{1}{2}\nabla \cdot \overline{\mathbf{u}(\mathbf{u}\cdot \mathbf{u})} = -\nabla \cdot \overline{p\mathbf{u}} - \frac{\overline{pw}}{r} - \overline{\mathbf{u}\mathbf{u}} \cddot \nabla \mathbf{U} 
-\frac{1}{2r} \left \{ 2Wq + \overline{w(\textbf{u}\cdot \textbf{u})} + 2\overline{u}^2W \right \} \\
+ \overline{uw}\frac{U}{r} +\nu \left \{ \mathbf{\mathcal{4}}q -\frac{(\overline{u^2}+\overline{w^2})}{r^2} + \frac{2}{r^2} (\overline{u\partial_\theta w} - \overline{w\partial_\theta u})\right \} - \nu\overline{\nabla \mathbf{u}\cddot \nabla \mathbf{u}}.
\end{gathered}
\end{equation}
We consider a statistically stationary flow that is homogeneous in the wall-parallel directions. Further, we assume that the net radial transport of TKE over the boundaries of a volume element in the turbulent BL is zero for $\delta_\nu \ll y \ll \delta$. We then arrive at a reduced form of (\ref{eq:tke}), where the net local production of TKE is equal to the local dissipation. 
\begin{equation}
\label{eq:redtke}
\overline{uw}\partial_r U - \frac{1}{r}\overline{uw}U = -\nu\overline{\nabla \mathbf{u}\cddot \nabla \mathbf{u}}.
\end{equation}
The first term on the left-hand-side of equation (\ref{eq:redtke}) represents the production of TKE due to a gradient of the mean streamwise velocity profile, i.e. shear. 
The curvilinear coordinate system gives rise to an additional production term (the second term), as compared to turbulent shear flow over a flat boundary. In fact, such additional production terms due to curvature appear both in the $u_\theta$-component equation and in the $w_r$-component equation, and are respectively, $\frac{1}{r}\overline{uw}U$ and $- \frac{2}{r}\overline{uw}U$. Together, they sum up to the second term on the left-hand-side in (\ref{eq:redtke}). 

The process of additional production of TKE by curvature of the streamlines may be explained by the conservation of angular momentum $\text{L}=Ur$ \citep{ray16b, tow56}. If one considers a vortex that exchanges two fluid elements from $r_1$ to $r_2$ where $r_1 < r_2$, the change in kinetic energy whilst conserving $\text{L}$ is $\Delta E_k = \frac{1}{2}(U_1^2r_1^2 - U_2^2r_2^2)(\frac{1}{r_1^2}-\frac{1}{r_2^2})$. For $(r_2-r_1)/r_1 \ll 1$, the change in $E_k$ can be rewritten as
\begin{equation}
\delta E_k = \frac{1}{r^3} \frac{d\text{L}^2}{dr^2}(\delta r)^2,
\end{equation}
where $\delta r \approx r_2-r_1$ and $r\approx r_1 \approx r_2$. This is a very similar energy exchange as for buoyancy stratified flows, where $\delta E = \beta g \frac{dT^2}{dz^2}(\delta z)^2$ \citep{tow76}. In fact, we see that if $d\text{L}/dr < 0$, the work carried out by the vortex is positive and the IC rotating and stationary OC TC flow might be called unstably stratified \citep{ray16b, ess96}, whereas for $d\text{L}/dr > 0$ (OC rotating, IC stationary) the vortex requires energy to survive and the flow is stably stratified.

\begin{figure}
\centering
\begin{subfigure}{.50\textwidth}
  \centering
  \includegraphics[width=1.0\linewidth]{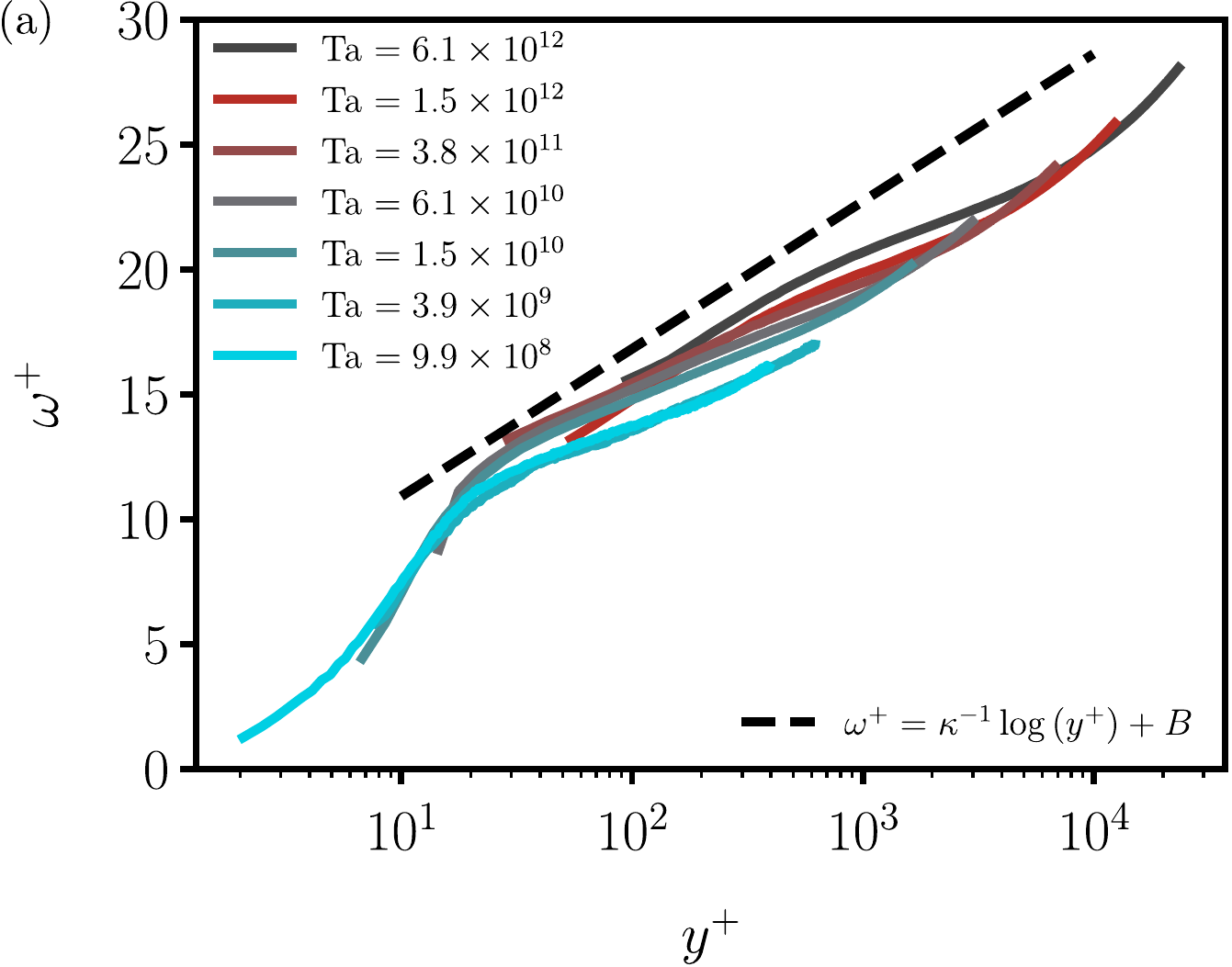}
\end{subfigure}%
\begin{subfigure}{.50\textwidth}
  \centering
  \includegraphics[width=1.0\linewidth]{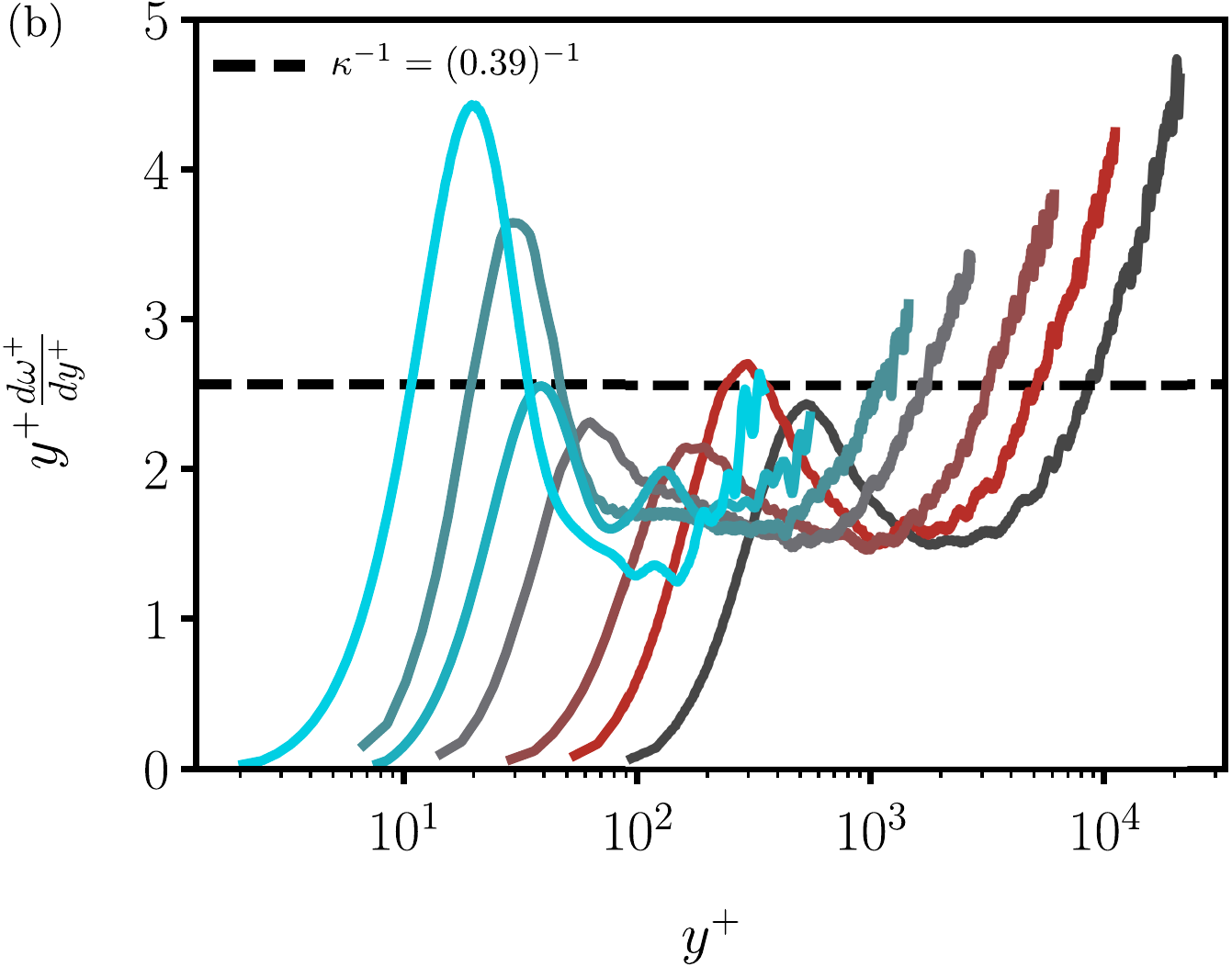}
\end{subfigure}%
\caption{The IC BL angular velocity profiles for $\eta=0.716$. (a) Mean angular velocity $\omega^+ = (\langle \omega(r)\rangle_{A(r),t}-\omega_i)/\omega_{\tau,i}$ versus the wall normal distance $y^+=(r-r_i)/\delta_{\nu,i}$. A logarithmic velocity profile with slope $\kappa^{-1}$ is observed in a limited spatial region at the highest Taylor numbers. (b) The diagnostic function reveals a very limited spatial region in which $y^+\frac{d\omega^+}{dy^+}=\kappa^{-1}$, indicated by the dashed line. Data from the PIV measurements of \cite{hui12}.}
\label{fig:omega}
\end{figure}

In pursuing this analogy, which we illustrate in figure \ref{fig:ill_ana}, we expect a region in the flow where ($\partial_rU \gg U/r$) from (\ref{eq:redtke}) such that the production of TKE is governed solely by shear, and the flow there behaves identical to flat plate BLs. Next to this, another region might exist where the production of TKE is governed solely by curvature effects ($U/r \gg \partial_rU$) and curvature stratification effects dominate. The demarcation line that separates the two regions is the location where both mechanisms are of comparable magnitude. \cite{bra69} recognized the similarity between buoyancy effects and streamline curvature, and derived the curvature analogy of the Obukhov length, here called $L_c$, with
\begin{equation}
\label{eq:mo_curv1}
\frac{L_c}{y} = \frac{\overline{uw}\partial_r U}{\frac{1}{r}\overline{uw}U},
\end{equation}
where $y=r-r_i$. We realize that in the overlap region the viscous stresses are negligible so that $\overline{uw}\approx u_\tau^2$ and the gradient of the streamwise velocity in the shear dominated region is $\partial_rU = \frac{u_\tau}{\kappa y}$, see (\ref{eq:vgrad}), which we take for reference in defining $L_c$. We approximate the curvature production by $U/r=\omega_i$, and $L_c$ then becomes
\begin{equation}
\label{eq:mo_curv2}
L_c = \frac{u_\tau}{\kappa \omega_i}.
\end{equation}
We use $\kappa=0.39$ throughout the paper, which is consistent with the data of \cite{hui12}, see figure \ref{fig:omega}, and also agrees with measurements of $\kappa$ in turbulent BLs and turbulent channel flows \citep{mar10}. However, we note that a range of $\kappa$ are reported in literature \citep{smi11}, and the employed data here is not conclusive on the second decimal. A subtle difference with the definition of \cite{bra69} resides in the definition of the curvature production term. \cite{bra69} uses the wall normal production only (i.e. $- \frac{2}{r}\overline{uw}U$), in strict analogy with the buoyancy production, that contains no streamwise production term. Here, however, we decide to use to sum of the streamwise and wall-normal curvature production terms (i.e. $- \frac{1}{r}\overline{uw}U$) to account for the total effects of streamline curvature. 

\subsection{Development of the functional form of $\omega^+(y^+)$}
\label{ssec:ICBL}
Figure \ref{fig:omega}(a) shows the angular velocity profiles for turbulent TC flow. For very high Re of $\text{O}(10^6)$, we observe the existence of a logarithmic form of the angular velocity profile with $\kappa\approx0.39$ and $B\approx 5$, in accordance with (\ref{eq:omgrad}).
However, the extent of the profile is very limited, namely $50<y^+<600$, as observed in \cite{hui12}, covering a much smaller spatial range than it would in canonical wall-turbulence systems \citep{pop00} at similar $Re_\tau$. Figure \ref{fig:omega}(b) presents the so-called diagnostic function, $y^+\frac{d\omega^+}{dy^+}$, which allows for a more detailed investigation of the slope of $\omega^+(y^+)$. Even for these high Re flows, only a very small region of the profile coincides with the straight line with slope $\kappa^{-1}$, which in this representation represents the log-layer. 

Following the analysis above, we expect the velocity profile to behave differently in the region where curvature effects play a role -- in close analogy with the Monin--Obukhov similarity theory. Hence, we make the wall-normal distance dimensionless with $L_c$, see (\ref{eq:mo_curv2}). This is done in figure \ref{fig:bulkprofiles}(b) where we plot the diagnostic function versus $y/L_c$. The result is a near perfect collapse of the angular velocity profiles, directly justifying the use of $L_c$ in turbulent TC flow. In fact, the profiles not only collapse with respect to their wall-normal location, but also in terms of their vertical coordinate, i.e. the slope of $\omega^+(y^+)$. This secondary flat regime with slope $\lambda^{-1}$ exists for larger $r>L_c$, than the $\kappa^{-1}$ regime. We find that $\lambda=0.64$. 

From these observations in figure \ref{fig:bulkprofiles} we obtain the unknown function $\Phi_\omega(\frac{y}{L_c})$ in (\ref{eq:vgrad_L}) for $0.20<y/L_c<0.65$:
\begin{equation}
\label{eq:lc_06}
\Phi_\omega \left( \frac{y}{L_c} \right) = \frac{1}{\lambda} \approx \frac{1}{0.64}; \quad \quad \quad 0.20\lesssim y/L_c \lesssim 0.65.
\end{equation}
Consequently, we integrate $\frac{d\omega^+}{d(y/L_c)}=\frac{1}{(y/L_c)\lambda}$ and arrive at
\begin{equation}
\omega^+ = \lambda^{-1}\log{(y/L_c)} + K,
\end{equation}
where $K$ is an integration constant and $\log$ is the natural logarithm. The offset $K$ of this second regime at larger $r$ is related to the height at which the first logarithmic regime at smaller $r$ peels off to the second log regime. We thus expect that $K = \kappa^{-1}\log{L_c^+} + C$ which results in,
\begin{equation}
\label{eq:om_fin}
\omega^+ = \lambda^{-1}\log{(y^+)} + (\kappa^{-1}-\lambda^{-1})\log(L_c^+) + C,
\end{equation}
where $C$ is a constant equal to $1.0$ (obtained by fitting to the highest Taylor number data). 
In figure \ref{fig:bulkprofiles}(a) we plot $\omega^+$ versus $y/L_c$ and subtract $K$ to highlight the collapse. Indeed, we observe a collapse of the profiles.

\subsection{The constant angular momentum region in the bulk}
\label{ssec:BULK}
\begin{figure}
\centering
\begin{subfigure}{.50\textwidth}
  \centering
  \includegraphics[width=1.0\linewidth]{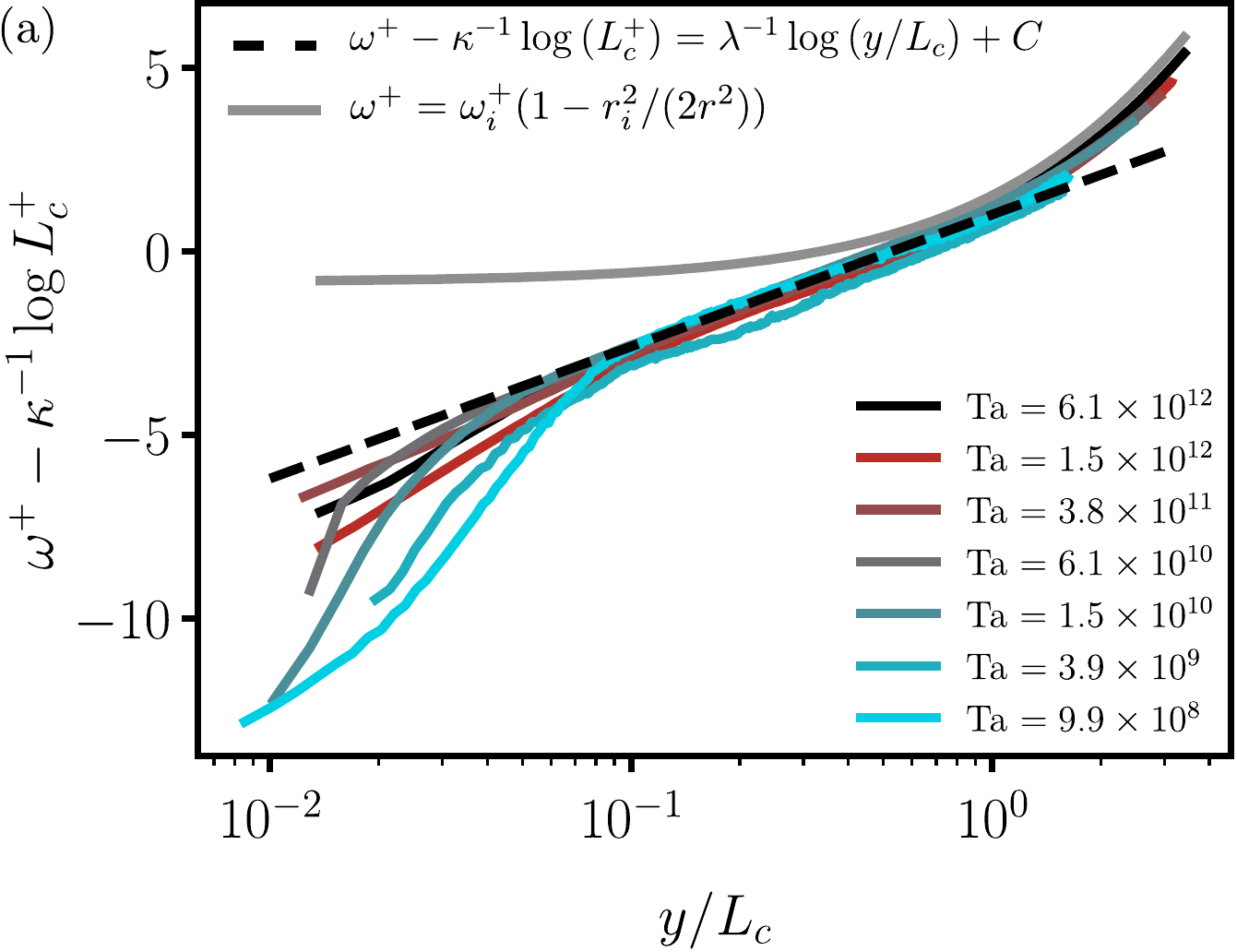}
\end{subfigure}%
\begin{subfigure}{.50\textwidth}
  \centering
  \includegraphics[width=1.0\linewidth]{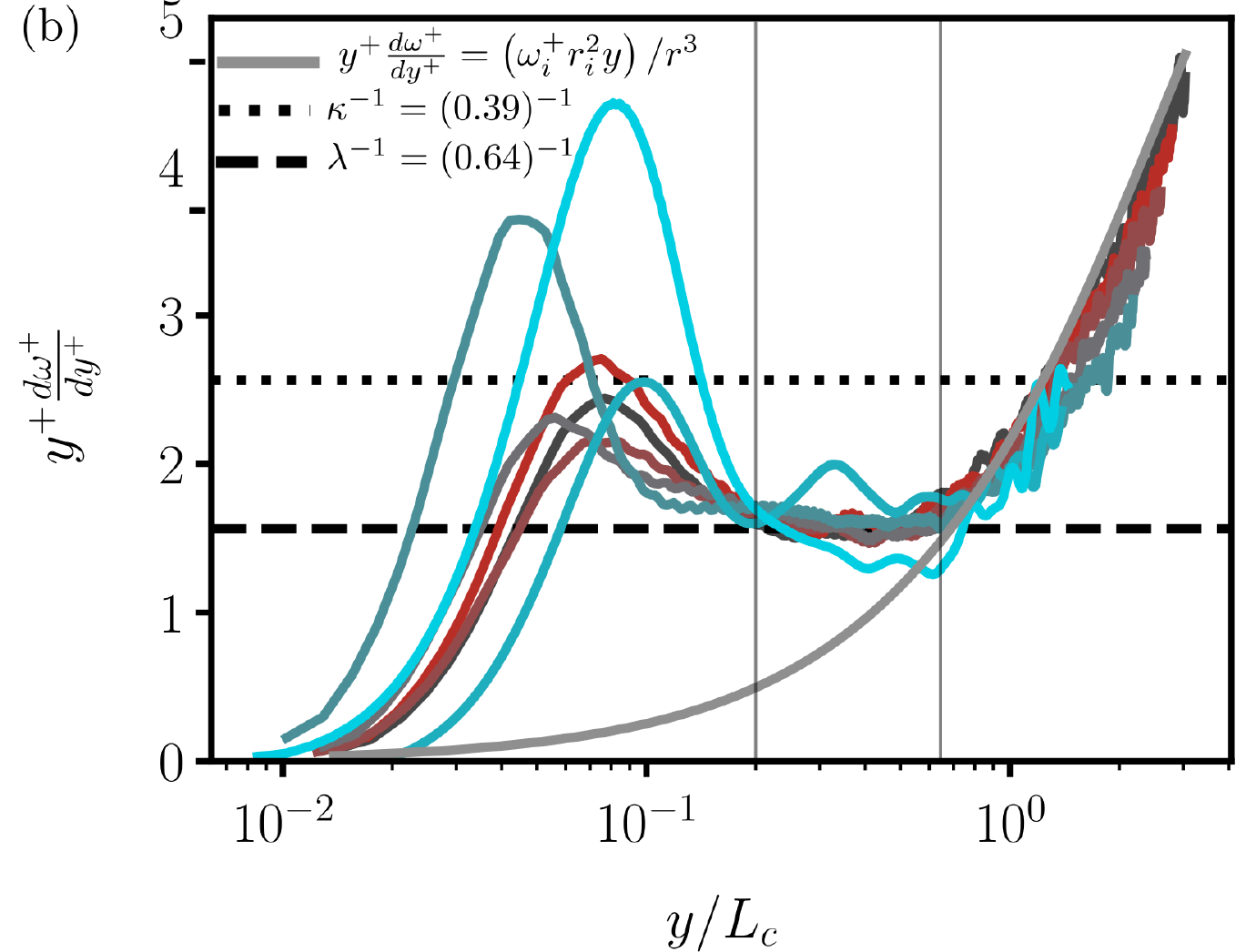}
\end{subfigure}%
\caption{The IC BL mean angular velocity profiles for $\eta=0.716$. (a) Mean angular velocity $\omega^+ = (\langle \omega(r)\rangle_{A(r),t}-\omega_i)/\omega_{\tau,i}$ with the $L_c^+$ dependent offset $\kappa^{-1}\log{(L_c^+)}$ subtracted to highlight collapse of the profiles. The curved, thick, grey line is the constant angular momentum $\text{M}_\text{o}=\omega_ir_i^2/2$, as derived by \cite{tow56}, which very closely fits the data at $y>L_c$. (b) Diagnostic function versus the rescaled wall normal distance $y/L_c=(r-r_i)/L_c$, where $L_c=u_{\tau,i}/ (\kappa \omega_i)$ is the curvature Obukhov length. The vertical grey lines indicate the bounds of the second log region. Data from the PIV measurements of \cite{hui12}.}
\label{fig:bulkprofiles}
\end{figure}
In the previous section we discussed the shape of the mean streamwise velocity profile in the IC BL, culminating in a new functional form which includes the stratification length $L_c$. However, to arrive at a Nu(Ta) relationship, we need to assess the velocity profile in the bulk region, too.
\cite{wen33} already observed that for unstable flows (i.e. IC rotation and a stationary OC) the bulk flow obeys a constant angular momentum $\text{L}=\text{M}_\text{o}$. Later, \cite{tow56} came to a similar conclusion and found that $\text{M}_\text{o} = \omega_ir_i^2/2$ for pure IC rotation. In recent years this finding is often confirmed by new datasets, see e.g. \cite{ost15b}, \cite{bra16}, and \cite{che19}.

Here, we plot the constant angular momentum region in figure \ref{fig:bulkprofiles}. We find that the transition from a $\lambda^{-1}$ region into a constant angular momentum $\omega^+ = \omega_i^+(1-r_i^2/(2r^2))$ region occurs at $y=L_c$. As such, the bulk region is entirely dominated by curvature effects of the streamlines. Consequently, the IC BL thickness $\delta_i$ is equal to the curvature Obukhov length, $\delta_i \approx L_c$ (and $\delta_o= 2.5 L_c$). Recently, a very similar thickness of the BL was empirically found by \cite{che19}.

\subsection{The outer cylinder boundary layer}
\label{ssec:OCBL}
\begin{figure}
\centering
\begin{subfigure}{.50\textwidth}
  \centering
  \includegraphics[width=1.0\linewidth]{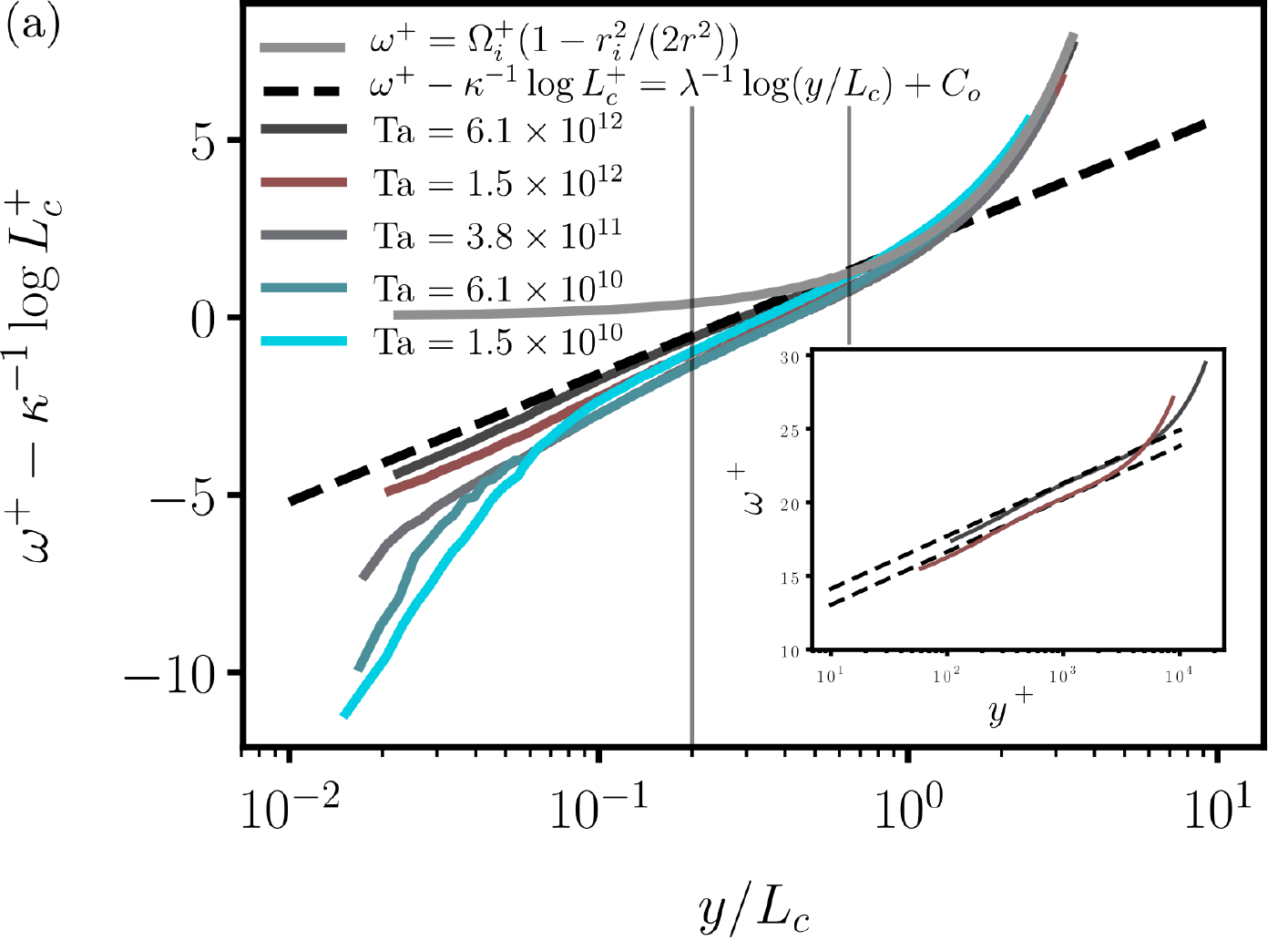}
\end{subfigure}%
\begin{subfigure}{.50\textwidth}
  \centering
  \includegraphics[width=1.0\linewidth]{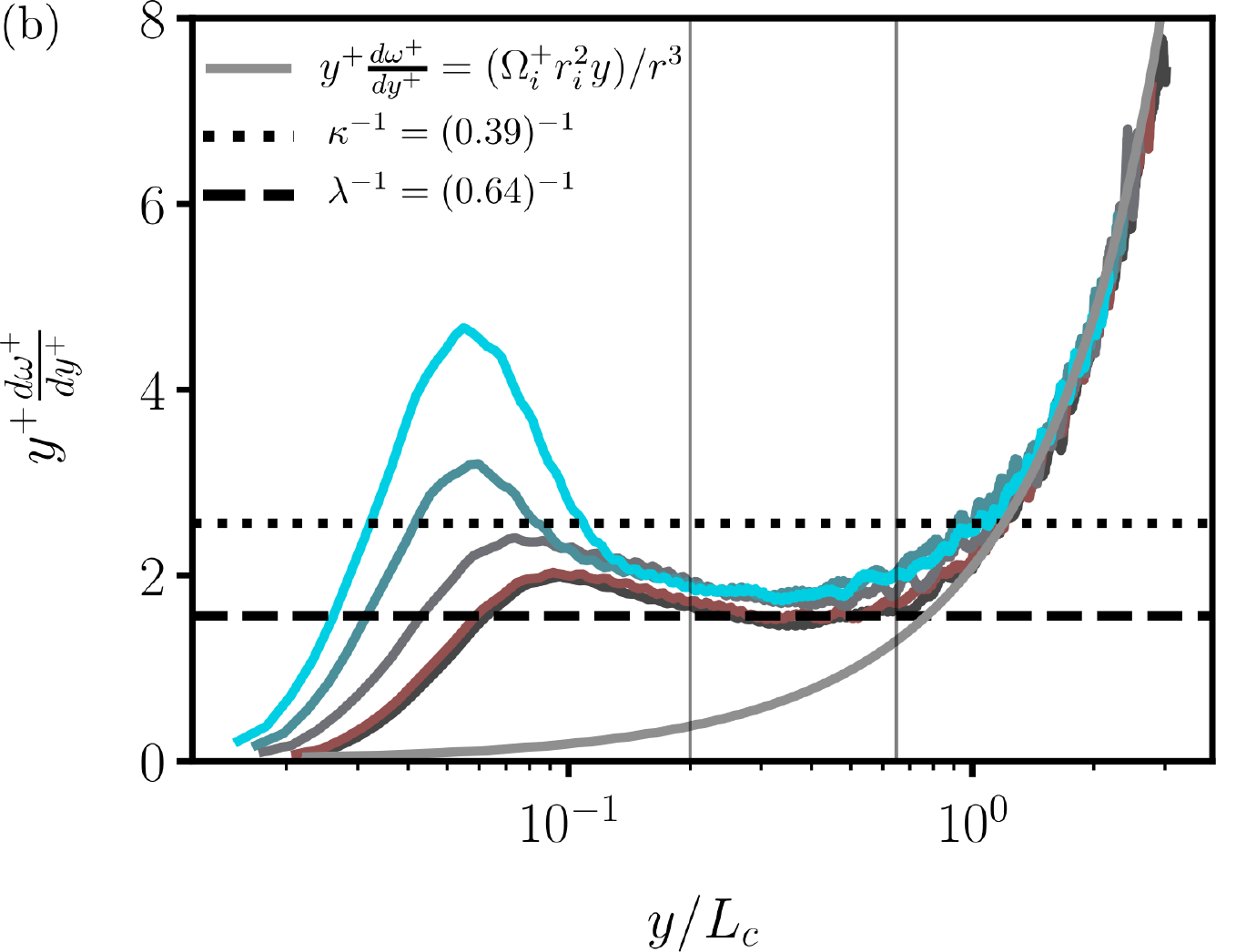}
\end{subfigure}%
\caption{The OC BL angular velocity profiles for $\eta=0.716$. (a) Mean angular velocity $\omega^+ = \langle \omega(r)\rangle_{A(r),t}/\omega_{\tau,o}$ with the $L_c^+$ dependent offset $\kappa^{-1}\log{L_c^+}+C_o$ subtracted to convey collapse of the profiles. The vertical grey lines indicate the bounds of the second log region. The curved, thick, grey line is the constant angular momentum $\text{M}_\text{o}=\omega_ir_i^2/2$, as derived by \cite{tow56}, which very closely fits the data at $y>L_c$. (b) Diagnostic function versus the rescaled wall normal distance $y/L_c=(r_o-r)/L_c$, where $L_c=u_{\tau,i}/ (\kappa \omega_i)$ is the curvature Obukhov length. For higher $r$ ($r>r_i+L_c$) the shear dominated logarithmic regime with slope $\kappa^{-1}$ peels off into a second logarithmic regime with slope $\lambda^{-1}$. The inset to (a) shows the mean angular velocity versus the wall normal distance $y^+=(r_o-r)/\delta_{\nu,o}$. Data from the PIV measurements of \cite{hui12}.}
\label{fig:huismanOCBL}
\end{figure}
Analogous to the IC BL we can analyse the OC BL in the spirit of the Monin--Obukhov similarity theory. As mentioned in \S \ref{sec:res1}, \cite{hui12} also obtained velocity profiles of the OC BL for the highest five Ta number experiments. 
From (\ref{eq:mo_curv1}) we derive that the relevant length scale for the OC BL is $L_{c,o} = r u_{\tau,o}/(\kappa U)$ with $y=r_o-r$. We approximate the velocity scale $U$ with $\omega_ir_i$ and the radius of curvature with $r_o$, so that $L_{c,o}=u_{\tau,i}/(\kappa \omega_i)$. The length scale is the same as $L_{c,i}$.  

Figure \ref{fig:huismanOCBL}(b) presents the gradient of the OC BL velocity profiles versus the dimensionless wall-distance $y/L_c$. Again, we observe collapse of the profiles in both the vertical direction and the horizontal direction. In the range $0.20<y/L_c<0.65$ the gradient of the profiles is $\lambda^{-1}$, whose value is identical to the IC BL profiles. 
Since the findings in figure \ref{fig:huismanOCBL}(b) are the same as in figure \ref{fig:bulkprofiles}(b), we derive the velocity profile for the OC BL in the same manner as (\ref{eq:lc_06}-\ref{eq:om_fin}) and arrive at 
\begin{equation}
\label{eq:om_fin)OC}
\omega^+_o = \lambda^{-1}\log{(y^+)} + (\kappa^{-1}-\lambda^{-1})\log(L_c^+) + C_o,
\end{equation}
where $C_o=2.0$ is obtained from fits in figure \ref{fig:huismanOCBL}(a). Again, the profiles in figure \ref{fig:huismanOCBL}(a) exhibit excellent overlap between (\ref{eq:om_fin)OC}) and the experimental data, especially at the highest two Ta numbers (see inset). We note that $Re_{\tau,o}$ at the OC BL is smaller than $Re_{\tau,i}$ at the IC BL, and consequently, we expect that the data at lower Ta still suffers from insufficient scale separation. 

We find that the obtained value for $C$ in (\ref{eq:om_fin}) differs from $C_o$ in (\ref{eq:om_fin)OC}). This is related to the different velocity scale in $L_c$ for the inner and outer cylinder BL. Once we estimate $L_{c,o}=u_{\tau,i}r_i/(\kappa U)$, where $U = 0.4\omega_i r_i$ is the angular velocity scale in the outer BL as obtained from the data ($Ta=6.1\times 10^{12}$), the constant $C_0=2-(\kappa^{-1}-\lambda^{-1})\log{0.4}=1.1 \approx C$ is consistent with (\ref{eq:om_fin}).

\section{The effects of the radius ratio $\eta$}
\label{sec:res4}
Up to this point, we have shown that one can treat inner cylinder rotating TC flow as an unstably stratified turbulent shear flow, in close analogy with temperature stratified flows. We proposed a new functional form of the mean angular velocity in (\ref{eq:om_fin}) that well describes the experimental profiles measured by \cite{hui12} in both inner and outer BL for all Re at $\eta=0.716$.
The question arises what implications of the theory of stratified flows -- and consequently (\ref{eq:om_fin}) -- bring to TC turbulence at varying radius ratios. To answer this question we first analyze DNS data of \cite{ost15b} and PIV data of \cite{vee16} at a lower radius ratio of $\eta=0.50$ (corresponding to larger curvature effects), followed by the analysis of the DNS \cite{ost15b} data at a high radius ratio of $\eta=0.91$. 

\subsection{Radius ratio $\eta=0.5$}
\begin{figure}
\centering
\begin{subfigure}{.50\textwidth}
  \centering
  \includegraphics[width=1.0\linewidth]{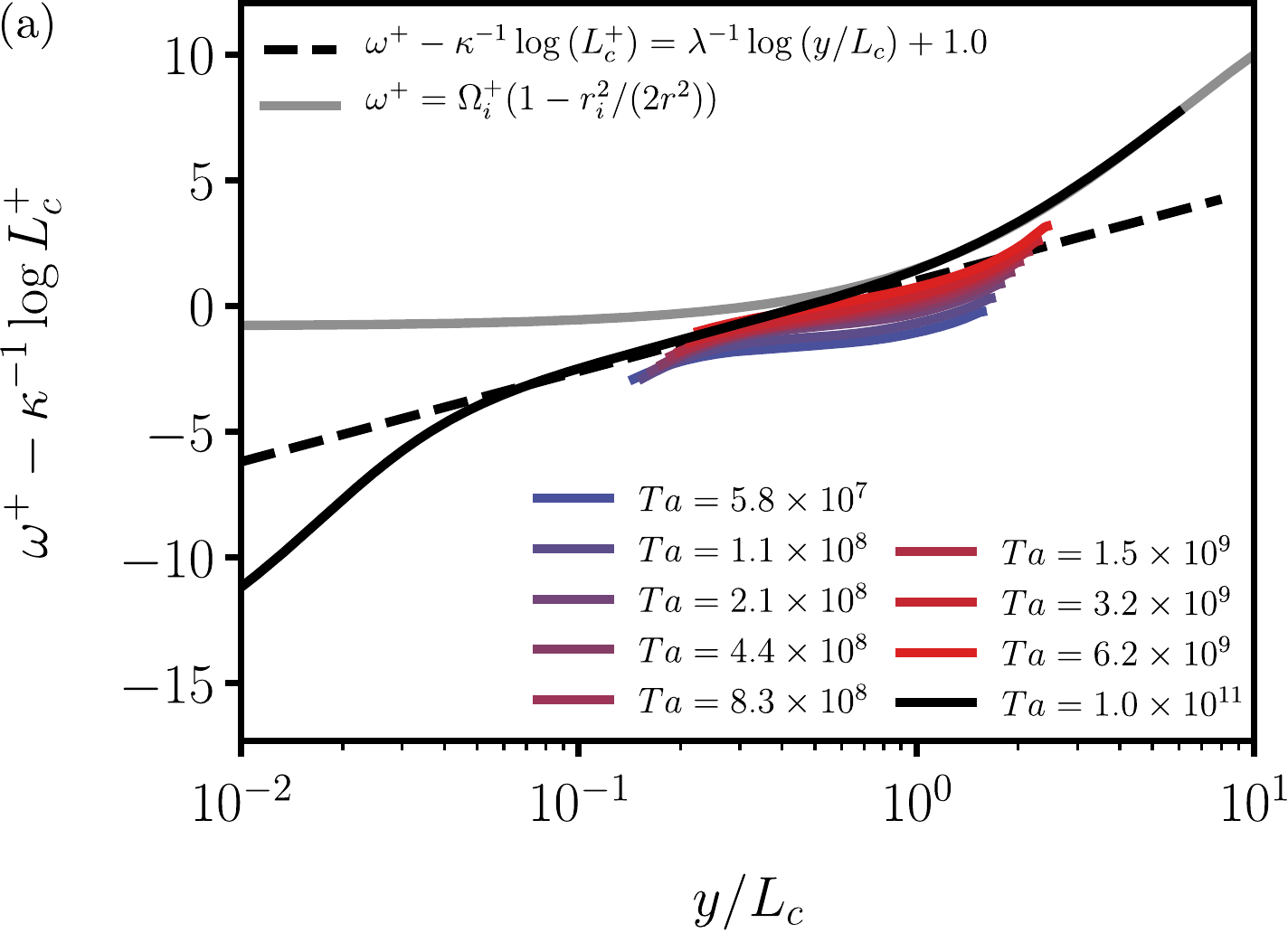}
\end{subfigure}%
\begin{subfigure}{.50\textwidth}
  \centering
  \includegraphics[width=0.96\linewidth]{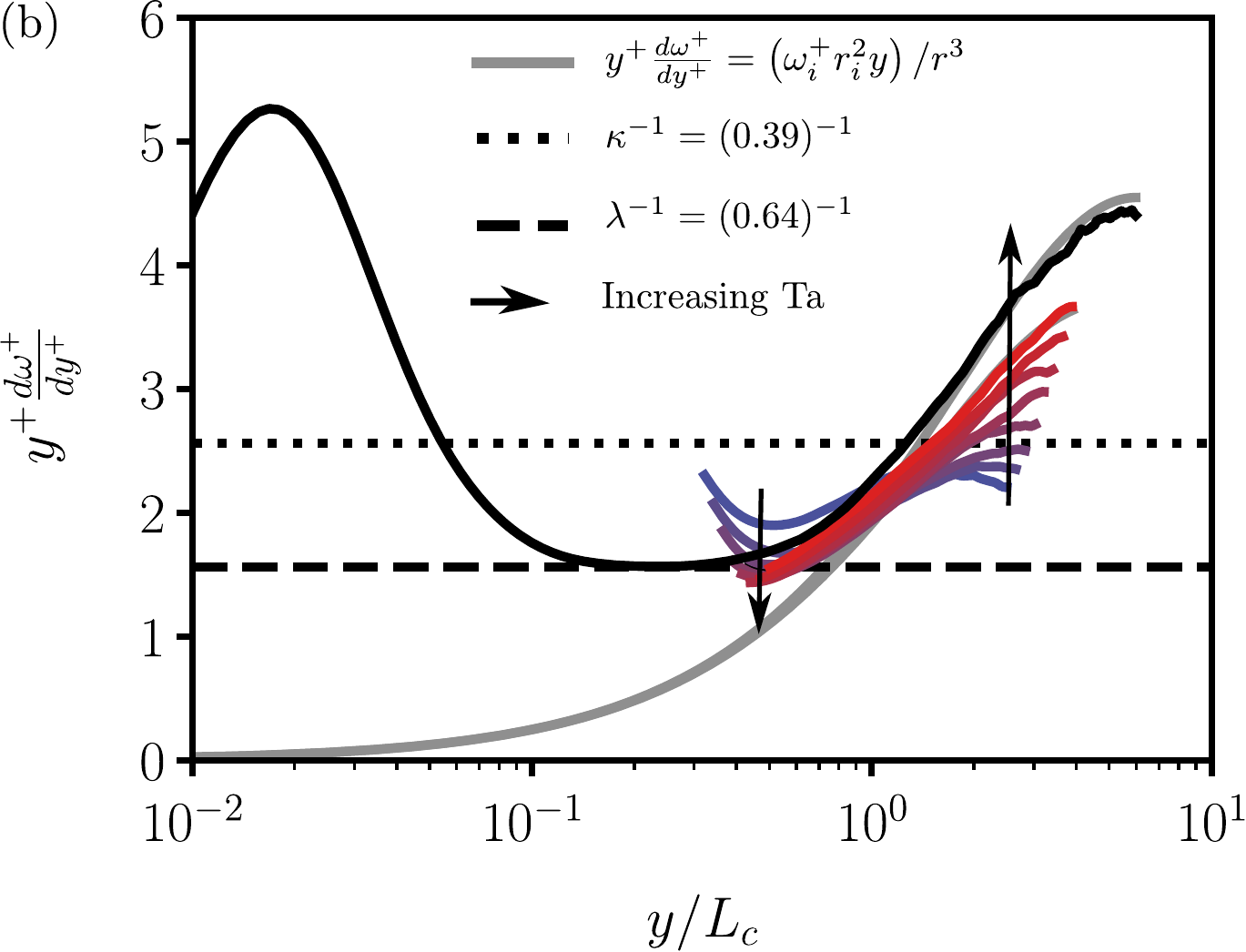}
\end{subfigure}%
\caption{The inner cylinder BL mean angular velocity profile at $\eta=0.50$. (a) Mean angular velocity $\omega^+ = (\langle \omega(r)\rangle_{A(r),t}-\omega_i)/\omega_{\tau,i}$ with the $L_c^+$ dependent offset $\kappa^{-1}\log{(L_c^+)}$ subtracted to convey collapse of the profiles. The curved, thick, grey line is the constant angular momentum $\text{M}_\text{o}=\omega_ir_i^2/2$, as derived by \cite{tow56}, which very closely fits the data at $y>L_c$. The black solid line represents DNS data of \cite{ost15b} whereas the colored lines represent the PIV data by \cite{vee16}. (b) Diagnostic function versus the rescaled wall normal distance $y/L_c=(r-r_i)/L_c$, where $L_c=u_{\tau,i}/ (\kappa \omega_i)$ is the curvature Obukhov length. }
\label{fig:eta05}
\end{figure}
Figure \ref{fig:eta05} presents the velocity profiles at $\eta=0.5$. The black solid line represents DNS data at a remarkable high Ta of $1.0\times 10^{11}$ resulting in a significant scale separation; $Re_\tau = 3257$, see table \ref{table_l}. Nevertheless, the diagnostic function in figure \ref{fig:eta05}(b) does not portray a shear dominated $\kappa^{-1}$ regime, i.e. the solid black line never follows the black dotted line. However, at $y/L_c\approx 0.20$ the $\lambda^{-1}$ regime is obtained. Note that we do not fit $\lambda^{-1}$ to the data, but only use the value ($\lambda=0.64$) as obtained in section \ref{sec:res1}. The dark grey solid line departs from the $\lambda^{-1}$ region around $y/L_c\approx 0.65$, to follow the $\text{M}_\text{o}=\omega_ir_i^2/2$ scaling of the bulk. This is in agreement with the observations at $\eta=0.716$. 

To understand the absence of a $\kappa^{-1}$ region for this low $\eta$, we refer to the scale separation in table $\ref{table_l}$. A $\kappa^{-1}$ slope requires that $1 \ll y^+ \ll 0.20L_c^+$. However, for $\eta=0.50$ at $\text{Ta }=1.0\times 10^{11}$ we find that $0.20L_c^+ \approx 100$, defying the existence of a shear production dominated region. The extensive scale separation between $L_c^+$ and $Re_\tau$ permits a large curvature dominated flow region where the angular momentum becomes constant, see figure \ref{fig:eta05}(a).
Figure \ref{fig:eta05} also presents in color the PIV data at low Ta. Although the scale separation is generally very low, with $Re_\tau$ not exceeding $10^{3}$, we find a trend towards the $\lambda^{-1}$ region with increasing Ta. Especially figure \ref{fig:eta05}(a) exhibits a collapse of the velocity profiles with (\ref{eq:om_fin}) at higher Ta. 

\subsection{Radius ratio $\eta=0.909$}
\begin{figure}
\centering
\begin{subfigure}{.50\textwidth}
  \centering
  \includegraphics[width=1.0\linewidth]{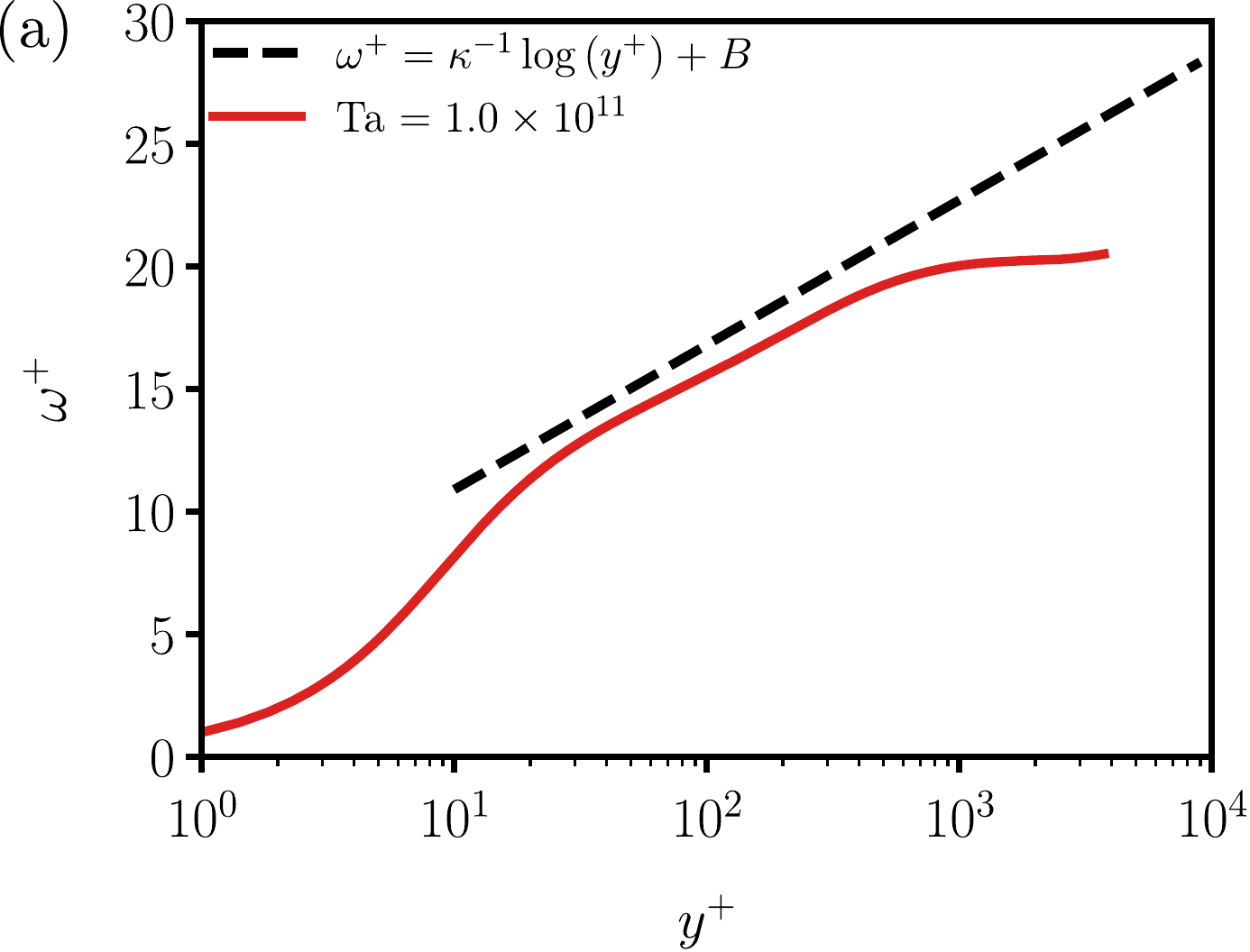}  
\end{subfigure}%
\begin{subfigure}{.50\textwidth}
  \centering
\includegraphics[width=1.0\linewidth]{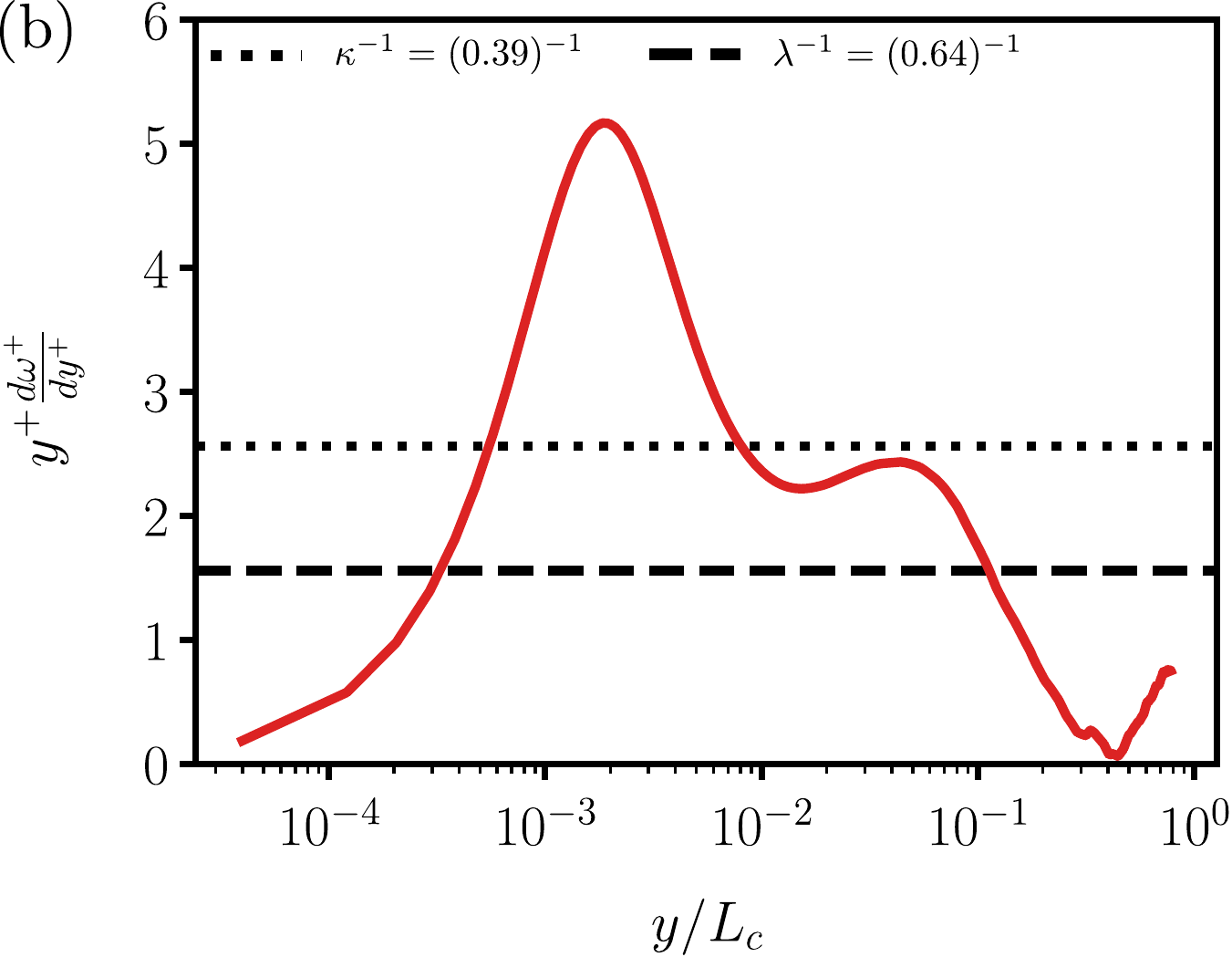}
\end{subfigure}%
\caption{(a) Mean angular velocity $\omega^+ = (\langle \omega(r)\rangle_{A(r),t}-\omega_i)/\omega_{\tau,i}$ versus the wall normal distance $y^+$. The red solid line is DNS data taken from \cite{ost15b}. (b) Diagnostic function versus the rescaled wall normal distance $y/L_c=(r-r_i)/L_c$, where $L_c=u_{\tau,i}/ (\kappa \omega_i)$ is the curvature Obukhov length, for $\eta=0.91$. }
\label{fig:eta09}
\end{figure}
Figure \ref{fig:eta09} shows data from a DNS at high $\eta=0.91$ (corresponding to small curvature effects) and $\text{Ta} = 1.0\times 10^{11}$. Interestingly, we observe a pronounced $\kappa^{-1}$ region. However there is a total absence of the $\lambda^{-1}$ and the $\text{M}_\text{o}$ region. Once again this is understood with the scale separation argument. In this case $L_c^+>Re_\tau$, and therefore there is no location in the flow where the curvature effects dominate, see table \ref{table_l}. 

\subsection{General radius ratio $\eta$}
\begin{figure}
\centering
\includegraphics[width=1.0\linewidth]{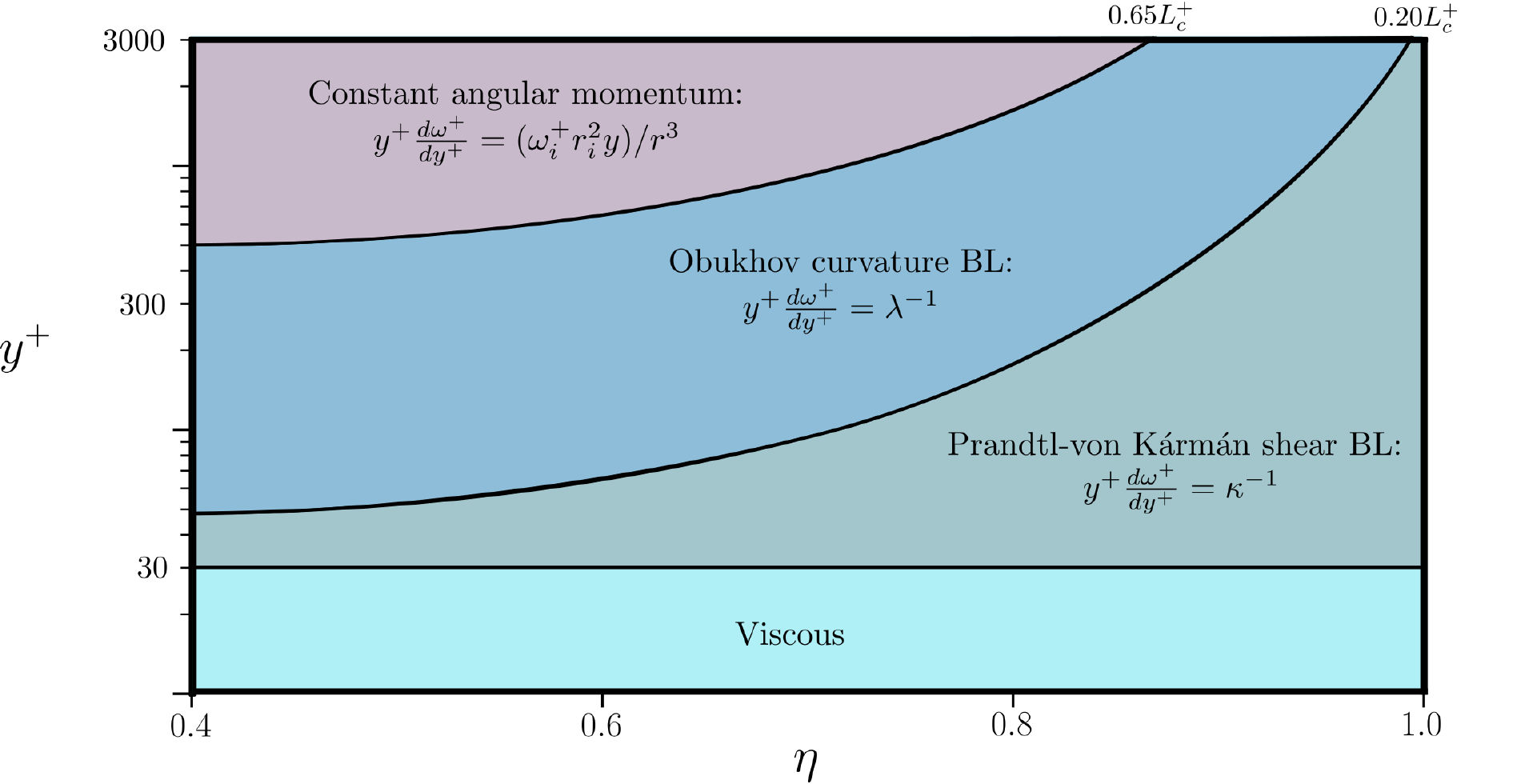}
\caption{Varying regimes in between the solid boundary (here the inner cylinder wall) and the outer length scale $Re_\tau$ for increasing radius ratio $\eta$, from $\eta=0.4$ (strong curvature) to $\eta=1.0$ (no curvature). The diagram is based on the values of $L_c^+$ at $Re_\tau\approx3000$ for $\eta=(0.500, 0.716, 0.909)$, see table \ref{table_l} in the appendix. }
\label{fig:phase_diagram}
\end{figure}  
To close this section, we provide a phase diagram of the scale separation at $Re_\tau\approx3000$ for varying $\eta$, in order to illustrate where one would expect to see $\kappa^{-1}$, $\lambda^{-1}$, and constant angular momentum regions of the angular velocity profile, in figure \ref{fig:phase_diagram}. We base the phase diagram on three cases for $\eta=(0.500, 0.716, 0.909)$ and $Re_\tau\approx 3000$, for which we calculate the phase boundaries, see table \ref{table_l}. Note that the boundaries are not sharp, and gradual changes in the relative importance of TKE production by shear and curvature lead to new regions. However, we now immediately see from the diagram that for high $\eta$ the Obukhov curvature BL is only expected to appear distinctly at extremely high $Re_\tau$ (higher than $Re_\tau=3000$). In contrast, for low $\eta$, we need extremely high $Re_\tau$ (higher than $Re_\tau=3000$) to observe the Prandtl-von K\'arm\'an turbulent BL type.

\section{The Nu(Ta) and $\text{C}_\text{f}(\text{Re}_\text{i})$ relationships}
\label{sec:res5}
\begin{figure}
\centering
\includegraphics[width=1.0\linewidth]{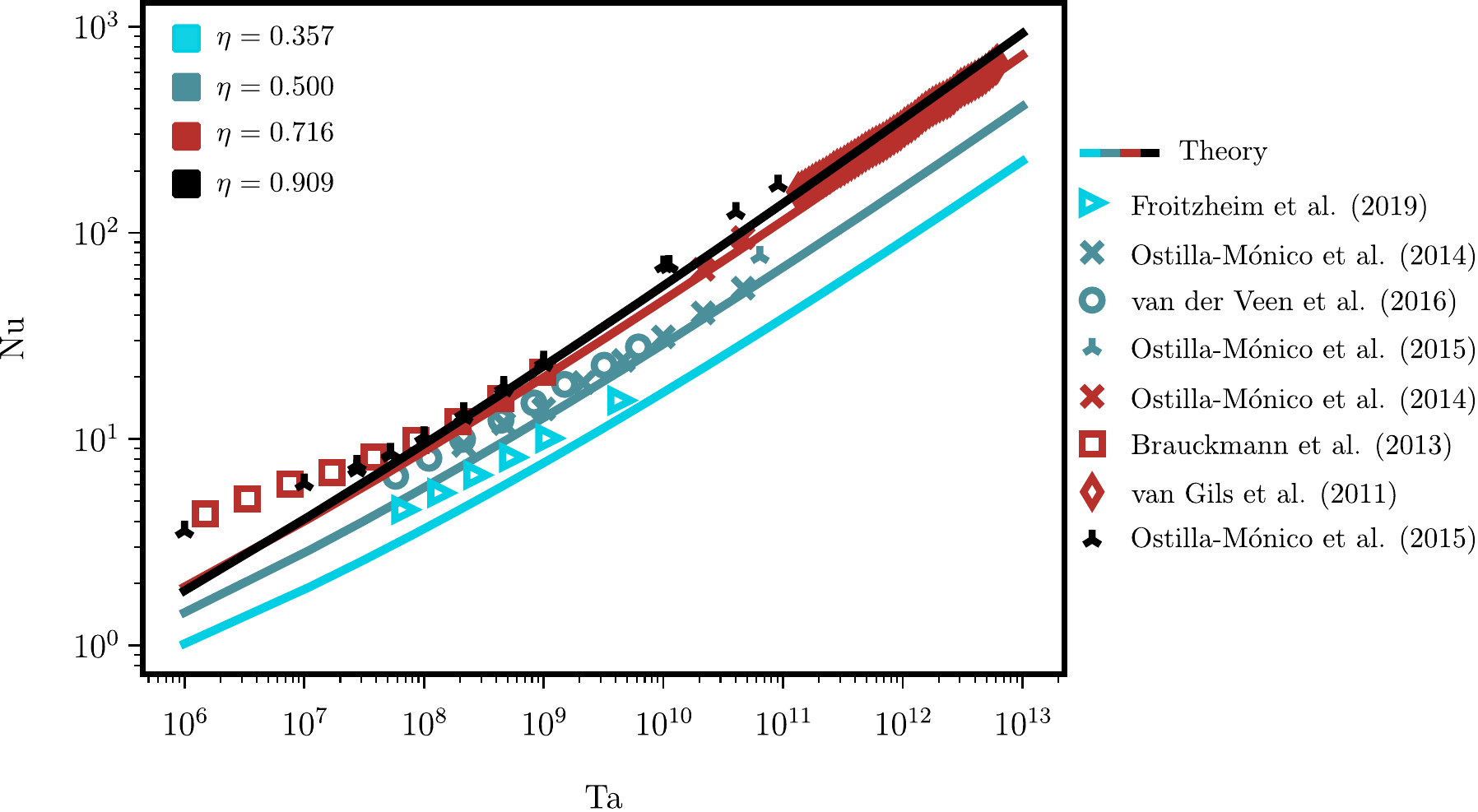}
\caption{The dimensionless torque Nu versus the dimensionless rotation rate Ta of the inner cylinder. Solid lines represent the theoretical prediction as derived by the matching of profiles in \S6, with the resulting relationship Nu(Ta) given by (\ref{eq:match3}). Symbols are the values of Nu obtained by DNS or experiments; $\eta=0.357$ (blue triangle) \cite{fro19}, $\eta=0.500$ (crosses) \cite{ost14d}, (open circles) \cite{vee16} and (triangle) \cite{ost15b}, $\eta=0.716$ (squares) \cite{bra13}, (crosses) \cite{ost14d} and (diamonds) \cite{gil11}, $\eta=0.909$ (triangles) \cite{ost15b}.}
\label{fig:nu_ta}
\end{figure}
The derivation of the angular velocity profile in a turbulent BL with strong curvature effects, see (\ref{eq:om_fin}), allows us to derive a functional form that relates the dimensionless torque Nu to the dimensionless driving Ta at $Re_o=0$. To do so, we follow the very recent work by \cite{che19}. Therein, the BL profile (the conventional shear dominated von K\'arm\'an type) is matched with the constant angular momentum bulk profile at the edge of the BL. With a fitting constant for the BL thickness, \cite{che19} arrive at a very accurate prediction of Nu over a wide range of Ta. Here, we match the angular velocity profiles in the bulk and the BL at the BL height $\delta = \alpha L_c$. Note that the constant $\alpha$ is easily extracted from figure \ref{fig:bulkprofiles}, where it refers to the outer bound of the $\lambda^{-1}$ region -- where the BL and bulk meet. 
\begin{equation}
\frac{\omega_i}{\omega_{\tau,i}} - \frac{1}{\lambda}\log \alpha L_c^+ - \left(\frac{1}{\kappa}-\frac{1}{\lambda}\right) \log L_c^+ - C = \frac{\omega_ir_i^2}{2\omega_{\tau,i}(r_i+\alpha L_c)^2}.
\label{eq:match1}
\end{equation}
We realize that $L_c=2r_iRe_\tau/(\kappa Re_i)$, $L_c^+=(4\eta Re_\tau^2)/(\kappa(1-\eta )Re_i)$, and $\omega_i/\omega_{\tau,i}=Re_i/(2Re_\tau)$ so that we can rewrite (\ref{eq:match1}),
\begin{equation}
\label{eq:match2}
\frac{Re_i}{2Re_\tau}- \frac{1}{\kappa}\log{\frac{4\eta Re_\tau^2}{\kappa(1-\eta )Re_i}} - \left(C + \frac{1}{\lambda}\log \alpha \right) - \frac{Re_i}{4Re_\tau \left( 1 + \frac{4\alpha Re_\tau}{\kappa Re_i} + \left(\frac{2\alpha Re_\tau}{Re_i}\right)^2\right)} = 0.
\end{equation}
This equation cannot be solved analytically. However, with the condition that $Re_\tau/Re_i \ll 1$, we can simplify the denominator in the last term of the left hand side to $1+\frac{4\alpha Re_\tau}{\kappa Re_i} + \left(\frac{2\alpha Re_\tau}{Re_i}\right)^2  \approx 1$ so that the equation can be solved \citep{che19}. We rewrite $\text{Nu}=2\eta(1+\eta)Re_\tau^2/Re_i$ and $\text{Ta}=(f(\eta)Re_i)^2$ with $f(\eta)$ defined in \S\ref{sec:tc}, and obtain
\begin{equation}
\label{eq:match3}
\text{Nu} = \frac{\kappa^2 \eta^3 \text{Ta}^{1/2}}{4(1+\eta)^2 W(Z)^2}, \quad \quad \quad Z= \sqrt{\frac{\kappa \eta^3 \text{Ta}^{1/2}}{2(1-\eta)(1+\eta)^3}}\exp{\left(\frac{\kappa \left( C+\frac{1}{\lambda}\log \alpha \right) }{2}\right)},
\end{equation}
where $W(Z)$ is the principal branch of the Lambert $W$ function. 

Figure \ref{fig:nu_ta} presents the prediction of equation (\ref{eq:match3}) together with 8 datasets from DNS and experiments -- covering $0.357 \le \eta \le 0.909$ and 7 orders of magnitude in Ta. $\alpha=0.65$, see (\ref{eq:lc_06}) and figure \ref{fig:bulkprofiles}. Naturally, we find deviations at low Ta, where the BLs are not fully turbulent yet. However, we find good overlap at high Ta for various $\eta$. For high $\eta$, (\ref{eq:match3}) looses its validity since shear is dominating curvature effects throughout the entire BL \textit{at the current Ta}. The Nu(Ta) relation is thus better described by the functional form derived in \cite{che19}. However, we note that the ratio $Re_\tau/L_c^+$ will become larger with increasing Ta, so that for extremely high Ta (even much higher than $10^{12}$), the Nu(Ta) relationship at $\eta=0.909$ will also follow (\ref{eq:match3}).

\begin{figure}
\centering
\begin{subfigure}{.50\textwidth}
  \centering
  \includegraphics[width=0.97\linewidth]{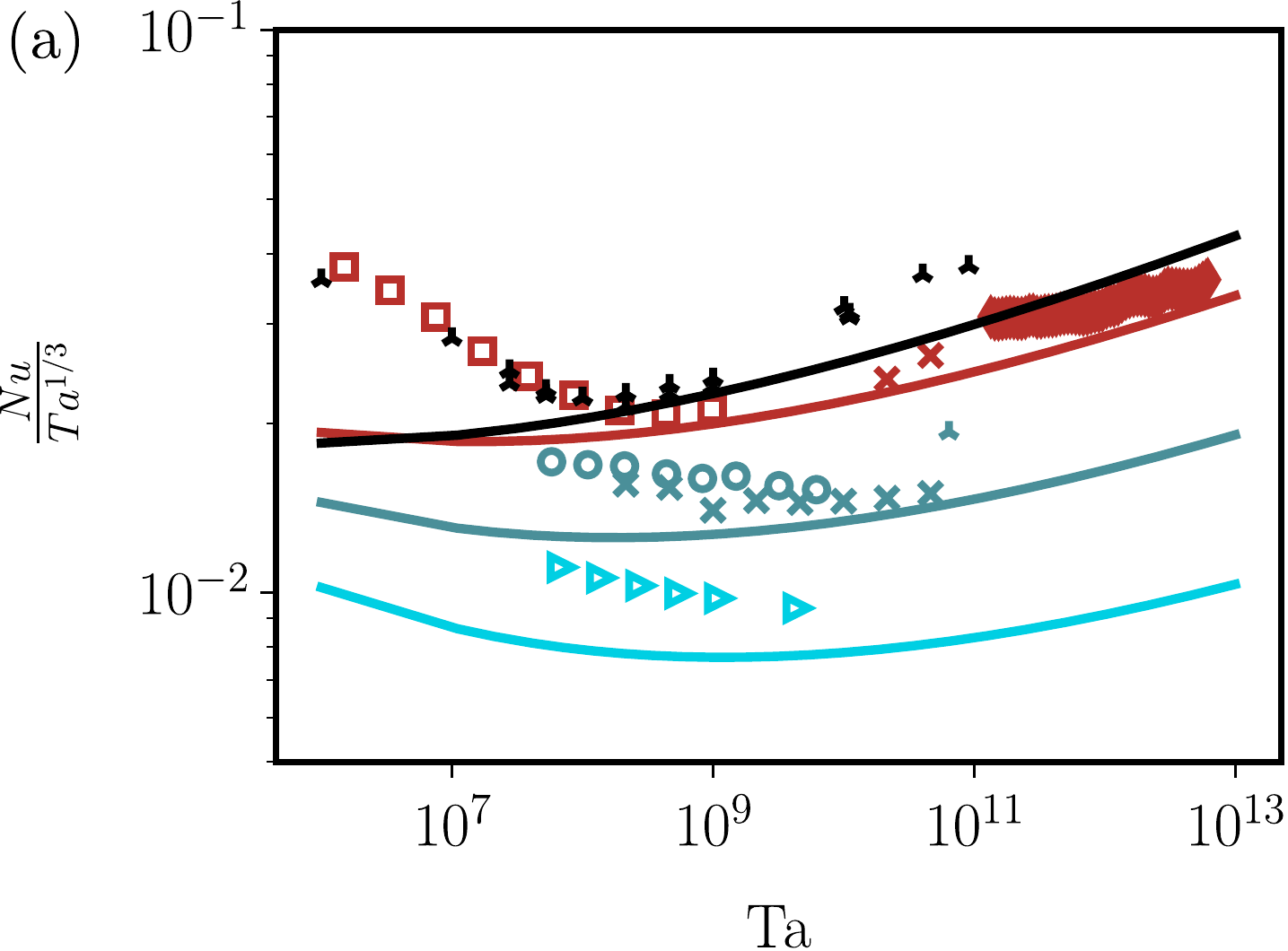}
\end{subfigure}%
\begin{subfigure}{.50\textwidth}
  \centering
  \includegraphics[width=0.95\linewidth]{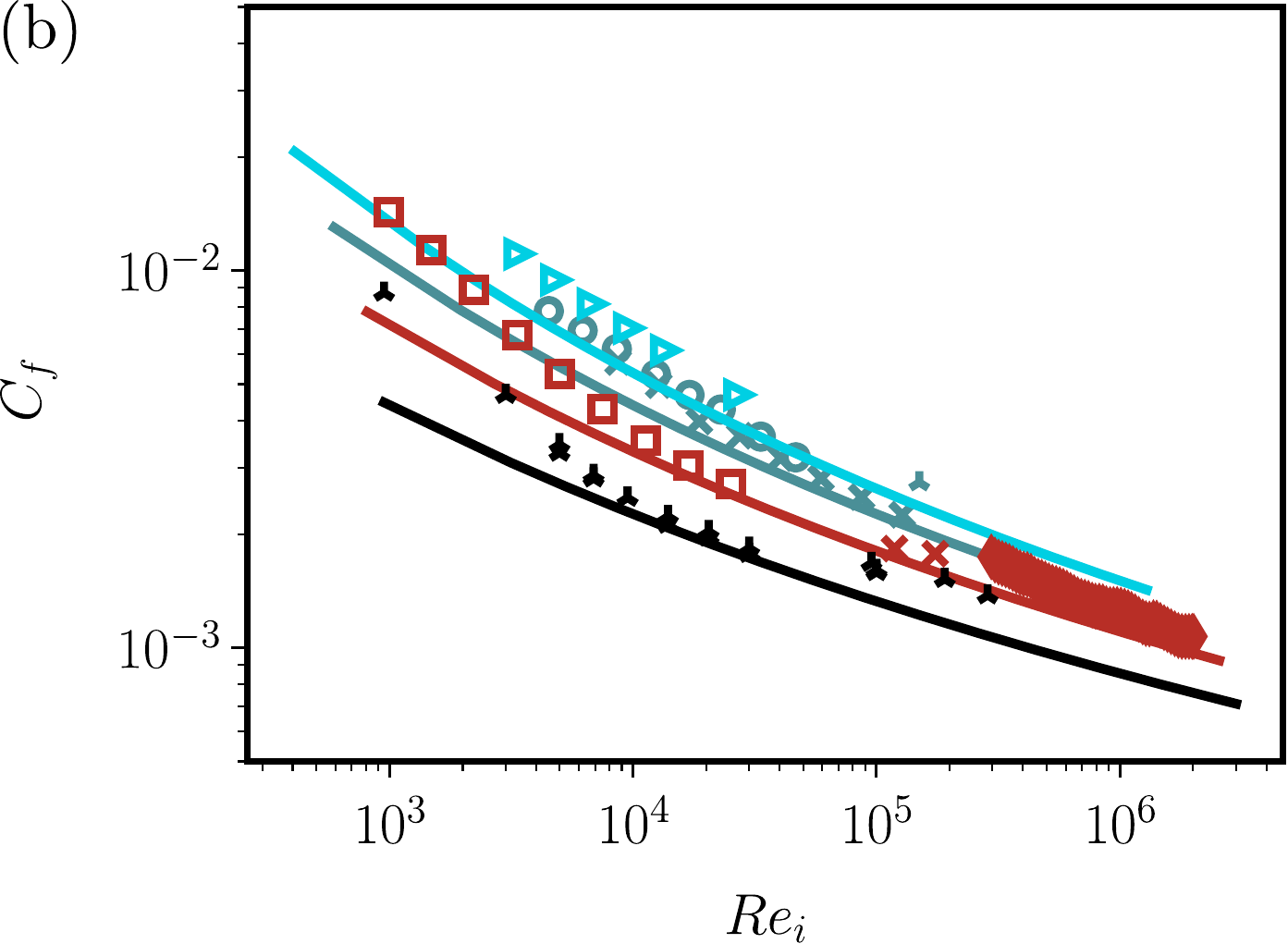}
\end{subfigure}%
\caption{(a) The dimensionless torque Nu, compensated with the scaling of TC flow with laminar BL and turbulent bulk $\text{Ta}^{1/3}$, versus the dimensionless rotation rate Ta of the inner cylinder. (b) The friction factor $C_f$ versus the the IC Reynolds number $\text{Re}_i$. Colours and symbols are the same as in figure \ref{fig:nu_ta} and links to the references can be found in the caption of that figure.}
\label{fig:comp_nu}
\end{figure}
For $\text{Ta}<10^6$, the BLs are of the laminar type and Nu scales with $\text{Ta}^{1/3}$ \citep{ost14}. Figure \ref{fig:comp_nu}(a) shows the Nu(Ta) relationship where Nu is compensated with $\text{Ta}^{1/3}$, such that we highlight the transition to a turbulent BL where the scaling exponent is larger than ${1/3}$. We emphasize that only after this transition, which is gradual and appears to depend on $\eta$, when BLs are entirely turbulent, equation (\ref{eq:match3}) will correctly calculate Nu(Ta). Figure \ref{fig:comp_nu}(b) presents the $\text{C}_f(\text{Re}_i$) diagram, which is more conventionally used in the pipe flow and BL flow communities. The solid lines are given by equation (\ref{eq:match3}) where the friction factor is calculated from $C_f = 4\text{Nu}/(\eta(1+\eta)\text{Re}_i)$.

\section{Summary and Conclusions}
In summary, we have developed a theory, similar to that of thermally stratified turbulent BLs, as famously developed by \cite{mon54}, for the curved turbulent BLs in inner cylinder rotating TC flow. In this analogy, the destabilizing effects from curvature of the streamlines in inner cylinder rotating TC flow are similar to the destabilizing effects coming from unstable thermal stratification in the atmospheric BL. 

We show that the curvature Obukhov length $L_c$ \citep{bra69} separates the spatial regions that are dominated by shear and curvature effects. We find that for $\delta_{\nu}<y \lesssim 0.20L_c$, the mean angular velocity profile in the BL is described by the classical shear profile, with the slope given by the von K\'arm\'an constant $\kappa^{-1}=0.39^{-1}$. In contrast, for $0.20L_c \lesssim y \lesssim 0.65L_c$, where curvature effects are relevant, the slope of the angular velocity profile is $\lambda^{-1}=0.64^{-1}$. For $y \gtrsim 0.65L_c$ curvature effects dominate, and a region with constant angular momentum sets in. This theory is applied to -- and found consistent with -- PIV measurements and high-fidelity DNS data covering a wide range of radius ratios $0.50 \le \eta \le 0.909$ and rotation rates $10^{8} \le \text{Ta} \le 10^{12}$, and describes both the IC BL and the OC BL. 

Building on these findings we derived a new functional form of the mean angular velocity profile in TC turbulence, with separate spatial regions where curvature and shear effects are respectively relevant. Upon matching \citep{che19} this BL profile with the constant angular momentum profile in the bulk, at the edge of the BL, we obtain a Nu(Ta) (and $C_f(Re_i)$) relation that agrees well with various data sets at high Ta and varying $\eta$.

Future research might investigate the effects of stably stratified TC flow (i.e. outer cylinder rotation), or even mixed stratified TC flow (i.e. counter cylinder rotation) within the framework of the Monin--Obukhov similarity theory. However, so far, only velocity profiles with a scale separation up to $Re_\tau\approx 1200$ are available for OC rotation \citep{ost16} to apply the theoretical analysis. Also, based on the newly derived velocity profile, it becomes necessary to reassess the fully rough asymptote for rough wall turbulent TC flow \citep{ber19}.

\section*{Acknowledgements}
This paper is devoted to Prof. Siegfried Grossmann on occasion of his 90th birthday. We congratulate him and thank him for all we learned from him on turbulence, physics, science, and beyond.

We are grateful to Sander Huisman, Rodolfo Ostilla-M\'onico and Roeland van der Veen for providing us their data. We further thank Ivan Marusic, Dominik Krug, Nicholas Hutchins and Myrthe Bruning for insightful discussions. This project is funded by the Priority Programme SPP 1881 Turbulent Superstructures of the Deutsche Forschungsgemeinschaft.

\clearpage
\section*{Appendix}
\label{appendix}
\begin{figure}
\centering
\begin{subfigure}{.50\textwidth}
  \centering
  \includegraphics[width=0.97\linewidth]{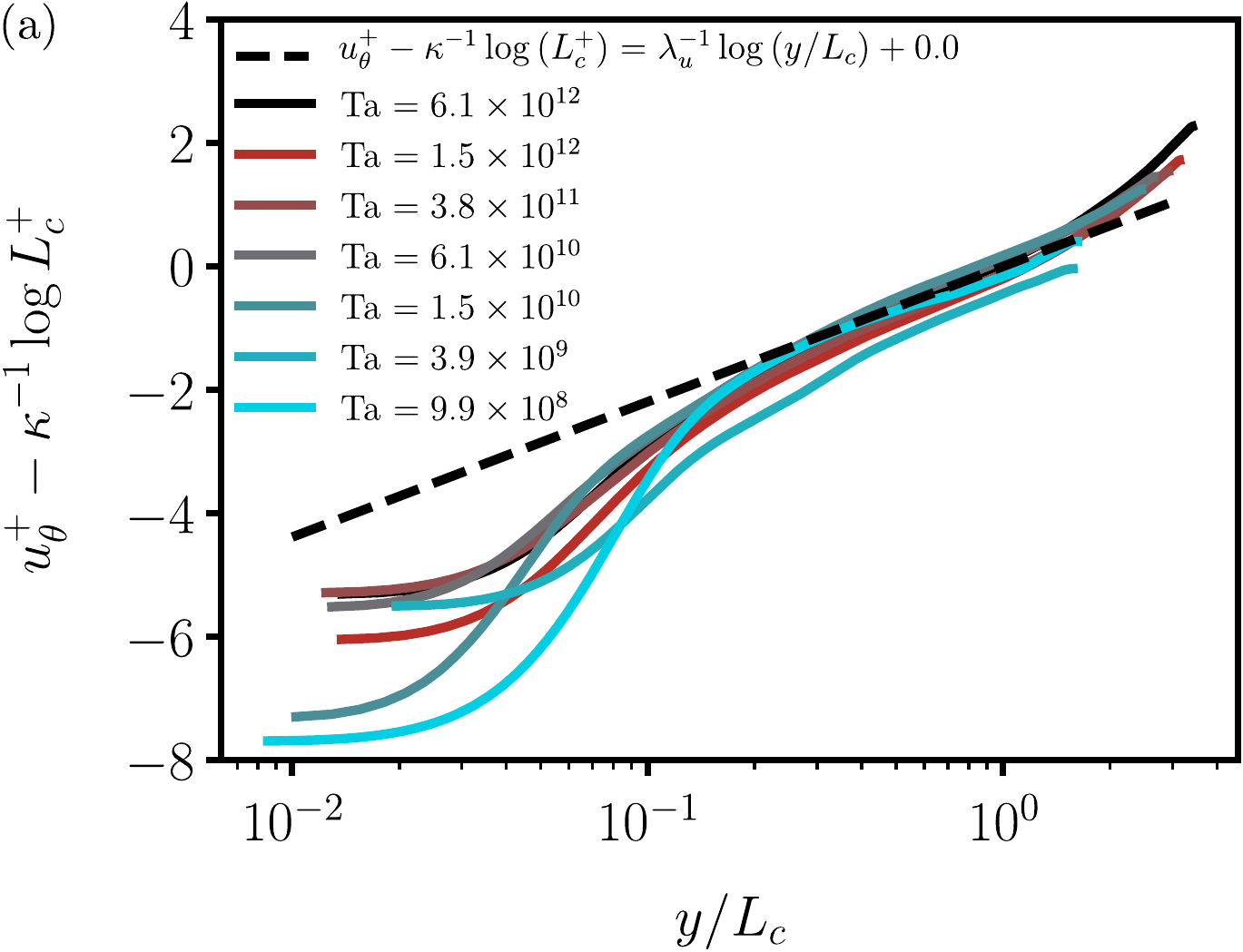}
\end{subfigure}%
\begin{subfigure}{.50\textwidth}
  \centering
  \includegraphics[width=0.97\linewidth]{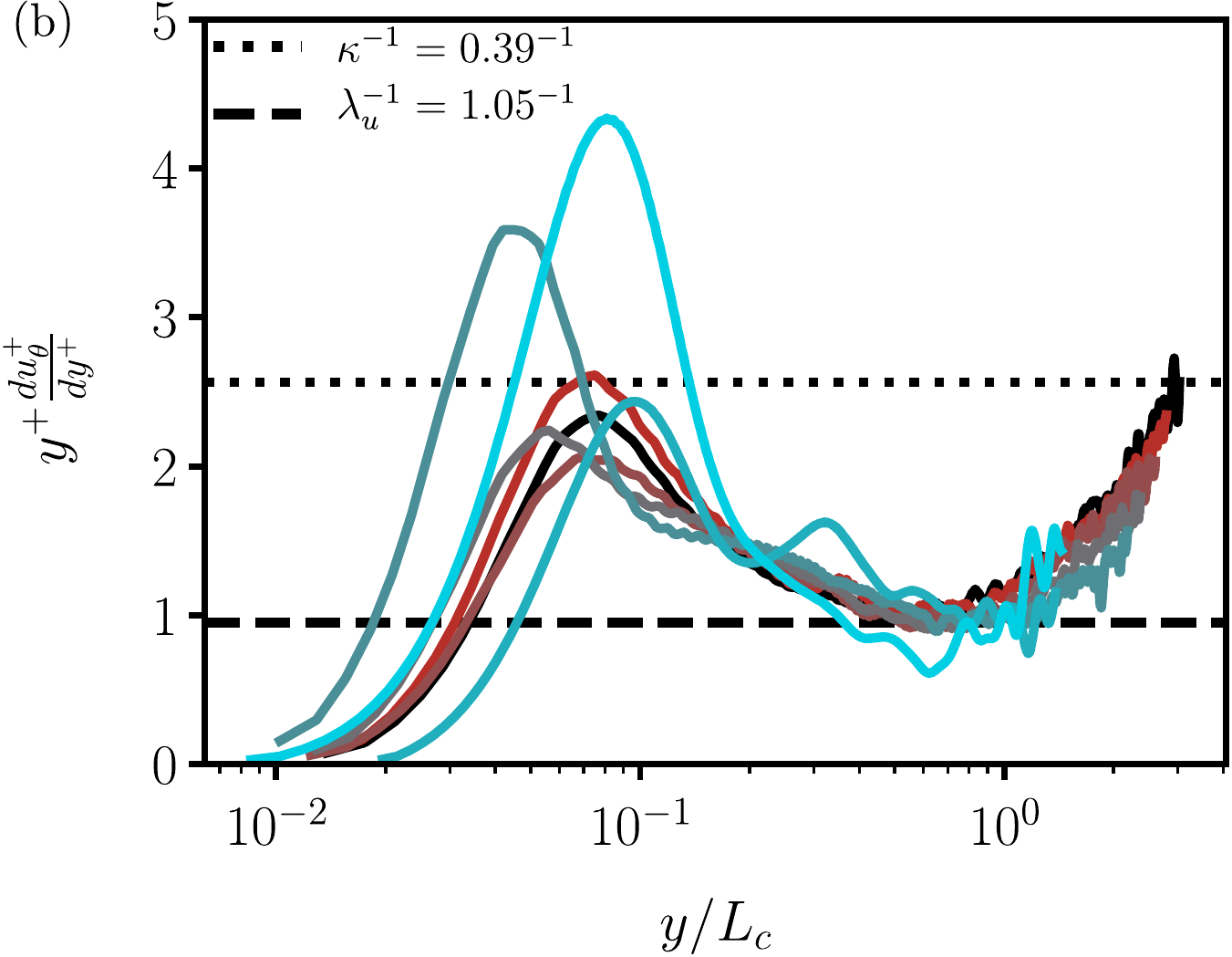}
\end{subfigure}%
\caption{The inner cylinder BL mean \textit{azimuthal} velocity profiles for $\eta=0.716$. (a) Mean azimuthal velocity $u_\theta^+ = (\langle u_\theta(r)\rangle_{A(r),t}-u_{\theta,i})/u_{\tau,i}$ with the $L_c^+$ dependent offset $\kappa^{-1}\log{(L_c^+)}$ subtracted to highlight collapse of the profiles. (b) Diagnostic function versus the rescaled wall normal distance $y/L_c=(r-r_i)/L_c$, where $L_c=u_{\tau,i}/ (\kappa \omega_i)$ is the curvature Obukhov length. Note that $\lambda_u^{-1}$ is different than $\lambda^{-1}$ in the main text. Data from the PIV measurements of \cite{hui12}.}
\label{fig:u_lc}
\end{figure}

See table 1 for an overview of the datasets.
\begin{table}
\centering
\begin{tabular}{c c c c c c c}
 $\eta$ & Ta  & $L_{c}^+$  & $Re_\tau$ &   \\ 
   \hline
    &  &       \cite{hui12}  &  &  \\
 0.716 & $9.9\times 10^8$  & $242$      & $488$ & PIV \\
 0.716 & $3.8\times 10^9$  & $395$    & $877$   & PIV \\
 0.716 & $1.5\times 10^{10}$ & $661$  & $1602$  & PIV \\
 0.716 & $6.1\times 10^{10}$ & $1124$  & $2950$  & PIV \\
 0.716 & $3.8\times 10^{11}$ & $2327$ & $6716$  & PIV \\
 0.716 & $1.5\times 10^{12}$ & $3947$ &$12217$  & PIV \\
 0.716 & $6.1\times 10^{12}$ & $6870$ &$23093$  & PIV \\  
 
 \hline
 
    &  &       \cite{vee16}  &  &  \\
  0.500 & $5.8 \times 10^{7}$  & $45$ & $141$ & PIV \\     
  0.500 & $1.1 \times 10^{8}$  & $55$ & $183$ & PIV \\     
  0.500 & $2.1 \times 10^{8}$  & $67$ & $239$ & PIV \\     
  0.500 & $4.4 \times 10^{8}$  & $84$ & $320$ & PIV \\
  0.500 & $8.3 \times 10^{8}$  & $103$ & $413$ & PIV \\          
  0.500 & $1.5 \times 10^{9}$  & $125$ & $531$ & PIV \\
  0.500 & $3.2 \times 10^{9}$  & $156$ & $714$ & PIV \\            
  0.500 & $6.2 \times 10^{9}$  & $192$ & $933$ & PIV \\       

 \hline

&  &       \cite{ost15b}    \\

 0.500 & $1.0\times 10^{11}$  & $544$ & $3257$  & DNS \\  
 0.909 & $1.0\times 10^{11}$  & $4794$ & $3745$ & DNS \\
    
\end{tabular}
\caption{Used datasets. The curvature Obukhov length $L_{c}^+$ and friction Reynolds number $Re_\tau$ at varying Ta and radius ratio $\eta$. } \label{table_l}
\end{table}

\clearpage
\bibliographystyle{jfm}
\bibliography{literature_turbulence}

\end{document}